\documentclass[twocolumn]{aastex701}
\usepackage[table]{xcolor}
\usepackage{adjustbox}
\usepackage{verbatim}
\usepackage{multirow}

\begin{document}

\title{A Bayesian Search for Planet Engulfment Signatures in Solar Analogs}

\author[orcid=0009-0006-4529-6724,gname=Zimo,sname=Cheng]{Zimo Cheng}
\affiliation{Department of Mechanical Engineering, Tsinghua University, Beijing 100084, China}
\email[show]{zimocheng1119@gmail.com}

% PI, supervisor, draft
\author[0000-0002-6937-9034]{Sharon Xuesong Wang}
\affiliation{Department of Astronomy, Tsinghua University, Beijing 100084, China}
\email[show]{sharonw@tsinghua.edu.cn}

% critical contribution
\author[0000-0003-4794-6074]{Fan Liu}
\affiliation{National Astronomical Observatories, Chinese Academy of Sciences, Beijing 100101, China}
\email{fanliu@bao.ac.cn}

% comments and discussion, formation of ideas
\author[0000-0001-8618-3343]{Haiyang S. Wang}
\affiliation{Center for Star and Planet Formation, Globe Institute, University of Copenhagen, Øester Voldgade 5-7, 1350 Copenhagen K, Denmark}
\email{haiyang.wang@sund.ku.dk}

% stellar sample, comments, discussion, collaboration project
\author[0000-0003-3281-6461]{Qinghui Sun}
\affiliation{Tsung-Dao Lee Institute, School of Physics and Astronomy, \& State Key Laboratory of Dark Matter Physics, Shanghai Jiao Tong University, Shanghai 201210, China}
\email{qinghuisun@sjtu.edu.cn}

% statistical framework
\author[0000-0003-0426-6634]{Johannes Buchner}
\affiliation{Max Planck Institute for Extraterrestrial Physics, Giessenbachstrasse 1, 85748 Garching, Germany}
\email{jbuchner@mpe.mpg.de}

% detailed comments, leading parallel paper
\author[0009-0006-7542-9639]{Mia Babatsikos}
\affiliation{School of Physics and Astronomy, Monash University, Clayton, Victoria 3800, Australia.}
\affiliation{OzGrav: Australian Research Council Centre of Excellence for Gravitational Wave Discovery, Clayton, VIC 3800, Australia.}
\email{Mia.Babatsikos@monash.edu}

% comments, discussion, collaboration project
\author[0009-0008-2988-2680]{Huiling Chen}
\affiliation{Department of Astronomy, School of Physics, Peking University, Beijing 100871, China}
\affiliation{Kavli Institute for Astronomy and Astrophysics, Peking University, Beijing 100871, China}
\email{chen-hl18@pku.edu.cn}

% stellar/disk sample
\author[0009-0002-5981-8520]{Jiayue Zhang}
\affiliation{Department of Astronomy, Tsinghua University, Beijing 100084, China}
\email{zhangjy23@mails.tsinghua.edu.cn}

% comments, discussion, collaboration project
\author[orcid=0000-0001-5082-9536]{Yuan-Sen~Ting} 
\affiliation{Department of Astronomy, The Ohio State University, Columbus, OH 43210, USA} 
\affiliation{Center for Cosmology and AstroParticle Physics (CCAPP), The Ohio State University, Columbus, OH 43210, USA}
\affiliation{Max-Planck-Institut f\"ur Astronomie, K\"onigstuhl 17, D-69117 Heidelberg, Germany}
\email{ting.74@osu.edu}

% comments, discussion, collaboration project
\author[0000-0003-0292-4832]{Zhen Guo}
\affiliation{Instituto de F{\'i}sica y Astronom{\'i}a, Universidad de Valpara{\'i}so, ave. Gran Breta{\~n}a, 1111, Casilla 5030, Valpara{\'i}so, Chile}
\affiliation{Chinese Academy of Sciences South America Center for Astronomy (CASSACA), National Astronomical Observatories, CAS,
Beijing 100101, China}
\email{zhen.guo@uv.cl}

% Zimo added Aaron's information with his permission.
\author[0000-0002-4442-5700]{Aaron Dotter}
\affiliation{Department of Physics and Astronomy, Dartmouth College, Hanover, NH 03755, USA}
\email{aaron.dotter@gmail.com}

% discussion
\author[0009-0004-9592-2311]{Serat M. Saad}
\affiliation{Department of Astronomy, The Ohio State University, Columbus, OH 43210, USA}
\email{saad.104@osu.edu}

% comments
\author[0009-0007-6769-7075]{Javier Osses}
\affiliation{Instituto de F{\'i}sica y Astronom{\'i}a, Universidad de Valpara{\'i}so, ave. Gran Breta{\~n}a, 1111, Casilla 5030, Valpara{\'i}so, Chile}
\email{javier.ossesp@postgrado.uv.cl}

\correspondingauthor{Zimo Cheng, Sharon X.~Wang}

\begin{comment}
List of authors and emails:
Fan Liu <fanliu@bao.ac.cn>, 
Mia Babatsikos <Mia.Babatsikos@monash.edu>
Ask Fan to reach out to Aaron Dotter

Qinghui Sun <qinghuisun@sjtu.edu.cn>
Haiyang Wang <haiyang.wang@sund.ku.dk>

Yuan-Sen Ting <ting.yuansen.astro@gmail.com>
Ask Yuan-Sen for the email of his student who was involved: saad.104@osu.edu

Megan Bedell <mbedell@flatironinstitute.org>
Johannes Buchner <johannes.buchner.acad@gmx.com>

Jiaxin Tang <tangjx22@mails.tsinghua.edu.cn>
Jiayue Zhang <zhangjy23@mails.tsinghua.edu.cn>
Huiling Chen <chen-hl18@pku.edu.cn>

Zhen Guo <zhen.guo@uv.cl>

\end{comment}
\begin{abstract}

We present a systematic Bayesian search for chemical fingerprints of planet engulfment in 113 solar twins and analogs with high-precision abundance measurements, 45 of which host known or candidate planets or brown-dwarf companions. We constructed a Bayesian framework with three sets of abundance models: random scatter, Galactic chemical evolution, and planet engulfment with bulk Earth or CM chondrite compositions. Through model comparisons, we identified three candidates whose abundance patterns strongly favor planet engulfment over the alternatives, with inferred engulfed masses of $\sim$7.5--33 $M_{\oplus}$. Our findings correspond to a nominal detection rate of $\sim$1--3\% for planet-engulfment signatures among solar analogs. This work extends abundance-based engulfment searches beyond the binary-star context and provides a framework for probing star–planet co-evolution with solar analogs, which goes beyond the commonly used abundance-condensation-temperature correlation ($T_{\text{c}}$ slope).

\end{abstract}



%%%%%%%%%%%%%%%%%%%%%%%%%%%%%%%%%%%%%%%%%%%%%%%%%%%%%%%%%%%%%%%%%%%%%%%%%%%%%%%%%
\section{Introduction}

Planets can be driven into their host stars by tidal decay or dynamical instability \citep[e.g.,][]{Levrard2009,Li2014,Yee2020}. If the accreted material is mixed into a sufficiently shallow stellar envelope, it can leave a measurable chemical imprint on the photosphere \citep[e.g.,][]{pinsonneault_mass_2001}. Most abundance-based searches specifically targeting planet engulfment have therefore focused on co-natal binaries, in which one component serves as a chemically matched control \citep[e.g.,][]{spina_gaia-eso_2015,oh_kronos_2018,behmard_planet_2023}. These studies have reported engulfment-signature fractions ranging from a few percent to several tens of percent \citep{spina_chemical_2021,liu_at_2024}, but their reliance on binary systems leaves open the question of whether comparable signatures can be identified in the much larger population of non-co-natal field stars, which is especially relevant for planet hosts.

Solar twins and analogs constitute a promising sample for searching for planet-engulfment signatures in single stars, as line-by-line differential analyses to the Sun can provide high-precision elemental abundance measurements \citep[e.g.,][]{bedell_abundance_2014}. Previous studies have identified that the Sun appears depleted in refractory elements compared with nearby solar twins and analogs \citep[e.g.,][]{melendez_peculiar_2009,bedell_chemical_2018}, which some hypothesize to originate from planet formation \citep[e.g.][]{chambers_slar_2010,booth_fingerprints_2020}, although there has been no observational evidence directly supporting such claims \citep[e.g.,][]{Gonzalez2010,Gonzalez2013}. These studies focused on the trends of elemental abundances with condensation temperature ($T_{\rm c}$ slope) or volatile-to-refractory ratios, which condense the multi-element abundance pattern into a single number and thus can be hard to interpret, particularly for non-co-natal stars, as Galactic chemical evolution (GCE), atomic diffusion, and other stochastic factors in star formation could also produce abundance variations \citep[e.g.,][]{adibekyan_origin_2014,nissen_high-precision_2015,huhn_how_2023}.

In this work, we developed a systematic Bayesian framework to model the full abundance pattern and identify planet-engulfment signatures among (non-co-natal) solar twins and analogs (Section~\ref{data}). Our work is motivated by \cite{liu_at_2024}, who found that the abundance pattern of engulfment signatures could be distinct compared with other astrophysical processes in co-natal pairs of stars. We thus also assume that, at least in some stars, planet-engulfment signatures could dominate the difference in abundance patterns between the target star and the Sun. Statistically, this means that a planet engulfment model, showing abundance patterns resembling those of bulk Earth or chondrites, would fit the data significantly better than alternative models, such as GCE or atomic diffusion (Section~\ref{method}). We identify three high-significance candidate stars with planet engulfment signatures (Section~\ref{result}) and discuss the implications on occurrence rate, comparison with inference using $T_{\rm c}$ slopes, and limitations and caveats of such a search (Section~\ref{discussion}).

%%%%%%%%%%%%%%%%%%%%%%%%%%%%%%%%%%%%%%%%%%%%%%%%%%%%%%%%%%%%%%%%%%%%%%%%%%%%%%%%%
\section{Data\label{data}}

Our sample consists of two sets of abundance measurements based on high-precision, line-by-line differential analyses: 79 solar twins and analogs from \citet{bedell_chemical_2018} and 42 from the Planets Around Solar Twins/Analogs (PASTA) survey \citep{sun_planets_2025,sun_planets_2025-1}. The PASTA sample comprises 40 stars with confirmed planets or high-probability TESS Objects of Interest, along with two non-planet-host stars also in the Bedell sample, observed for calibration and comparison purposes. In addition, there are 6 planet-hosting solar twins/analogs in both samples, thus a total of 8 overlapping stars. We prioritize these two datasets over some of the other large samples available in the literature, for example, the large catalog of abundances derived from machine-learning algorithms in \citet{rampalli_sun_2024} and \cite{martos_signatures_2025}, because line-by-line differential analyses starting from hands-on measurements of equivalent widths provide the highest precision available in the context of this work \citep{bedell_abundance_2014}.

We model the abundances of 19 elements from C through Zn and exclude elements with atomic numbers $Z>30$, whose abundances were derived through spectral synthesis rather than the equivalent-width procedures used for the rest of the elements. The Bedell and PASTA samples share the same set of elements, except for potassium (K), which is unavailable in the Bedell sample. We updated the companion information for each star by querying SIMBAD and the NASA Exoplanet Archive. In total, we have 45 stars hosting planets or brown dwarfs, including 11 stars in the Bedell sample and 40 in the PASTA sample (6 overlap). In summary, our dataset contains 113 unique stars with 121 sets of survey-specific abundance measurements (8 overlap). Appendix~\ref{app:data} details the construction of our sample.

According to the abundance comparison for the 8 overlapping stars in the Bedell and PASTA samples, the consistency between the two samples is insufficient to support the combination of the two samples due to differences in the instruments, spectral resolution, and SNR of the spectra in each sample \citep[Figure 1,][]{sun_planets_2025-1}. Correcting for instrumental offsets and potential systematics is being investigated in other ongoing PASTA collaboration projects. Since modeling and model comparisons are performed for each star on an individual basis, this inconsistency would not affect our main conclusion, with the exception of the engulfment occurrence rate estimate (see Section~\ref{result} and \ref{diss:rate}). The eight overlapping targets are counted only once when reporting the number of unique stars and serve as a validation test of the effects of such systematic inconsistencies.

%%%%%%%%%%%%%%%%%%%%%%%%%%%%%%%%%%%%%%%%%%%%%%%%%%%%%%%%%%%%%%%%%%%%%%%%%%%%%%%%
\section{Methods}
\label{method}

We search for stars whose abundance patterns are better described by planet engulfment than by plausible non-engulfment alternatives. For each star, we model the abundance vector relative to either the Sun or the mean of the corresponding sample. Our approach does not attempt to fit all physical processes simultaneously. Instead, we work with the basic assumption that, at least for some stars, there may be a dominant factor that shapes the abundance pattern. We test whether an engulfment template provides a better dominant description of the observed abundance pattern than each of the alternative models considered below.

\subsection{Abundance-pattern models}
\label{main_models}

We consider three classes of models: a flat abundance pattern, Galactic chemical evolution (GCE), and planet engulfment. Every model includes an intrinsic-scatter parameter, $\sigma_{\rm int}$, which is added in quadrature to the reported abundance uncertainties (see Equation \ref{eqn:likelyhood}).

The flat model, $F$, consists of an element-independent offset plus intrinsic scatter. It represents the null hypothesis that the abundance vector contains no resolved element-dependent structure. Atomic diffusion can also alter photospheric abundances, but the currently available diffusion calculations do not cover the full set of elements used here. Moreover, differential diffusion effects between elements are substantially reduced when using [X/Fe] instead of [X/H] for the narrow range of stellar parameters in our sample \citep{dotter_influence_2017,moedas_atomic_2022}. We therefore adopt [X/Fe] for our fiducial analysis. The details and limitations of this approximation are discussed in Appendix~\ref{atomic_diffusion}.

We tested two GCE models. The first, $G_{\rm Bedell}$, assumes linear [X/Fe]-age relations. We adopted the best-fit GCE slopes from \citet{bedell_chemical_2018} for the Bedell sample and from \citet{sun_planets_2025} for PASTA. The second, $G_{\rm GALAH}$, is a two-process model calibrated using GALAH DR3 \citep{griffith_abundance_2019,weinberg_chemical_2019,griffith_residual_2022,weinberg_chemical_2022}. It combines prompt chemical enrichment from massive stars and core-collapse supernovae with delayed enrichment from Type~Ia supernovae. Because two-process coefficients are available for only a subset of the measured elements, each set of model comparisons involving $G_{\rm GALAH}$ uses the same set of elements for the engulfment model. The details for these two GCE models are given in Appendix~\ref{gce}.

The engulfment model, $E$, follows the model commonly adopted in previous works \citep{chambers_slar_2010,oh_kronos_2018,behmard_planet_2023,liu_at_2024}. It assumes that engulfed material is well mixed throughout the stellar convection zone. For element $X$, the predicted enhancement is

\begin{equation}
E_{\text{[X/H]}}(M_{\rm p},f_{\text{cz}}) = \log_{10} \frac{M_{\rm p}f_{X,\text{planet}} + M_{\text{star}} f_{\text{cz}}f_{X,\odot}}{M_{\text{star}} f_{\text{cz}}f_{X,\odot}},\label{eqn:raw_engulf}
\end{equation}
where \(M_{\rm p}\) is the engulfed mass, \(f_{\text{cz}}\) is the mass fraction of the star's convection zone, and $f_{X,\rm planet}$ and $f_{X,\odot}$ are the mass fractions of element $X$ in the engulfed material and the convection zone of the baseline star, respectively. For [X/Fe], we subtract the corresponding predicted Fe enhancement. 

Equation~\ref{eqn:raw_engulf} shows a degeneracy between the engulfed mass and the stellar convection-zone mass. We thus fitted the equivalent engulfed mass, defined as

\begin{equation}
M_{\rm p,eq} =
M_{\rm p}
\frac{M_\odot f_{{\rm cz},\odot}}
{M_\star f_{\rm cz}}.
\label{eqn:meq}
\end{equation}

We test two compositions for the engulfed materials: bulk Earth \citep{allegre_chemical_2001} and CM chondrite \citep{wasson_compositions_1988}. For the [X/H]-based tests, we also consider an engulfment model with a constant offset to account for an initial metallicity difference between the star and the adopted baseline (the Sun or the sample mean). Full model expressions and parameter priors are provided in Appendix~\ref{app:expression} and \ref{app:prior}.

\subsection{Bayesian model comparison}

For each star, we fit the models using a Gaussian likelihood that includes both the reported abundance uncertainties and $\sigma_{\rm int}$ (see Appendix \ref{app:bayesian_method}). We calculate the Bayesian evidence, $Z$, with \textsc{dynesty} \citep{speagle_dynesty_2020,sergey_koposov_2025_17268284,skilling_nested_2004,skilling_nested_2006,feroz_multinest_2009}, using 500 live points and a stopping criterion of 0.01. We quantify the preference for engulfment over a null model \(N\) through the log Bayes factor

\begin{equation}
\Delta\ln Z(E-N) \equiv \ln Z_E-\ln Z_N,
\end{equation}

where $N$ is $F$, $G_{\rm Bedell}$, or $G_{\rm GALAH}$. Model pairs are always evaluated over identical element sets so that their evidence differences are not driven by unequal numbers of abundance measurements.

We perform the analysis using either [X/H] or [X/Fe], and with either the Sun or the survey sample mean as the abundance baseline. These alternative setups test the sensitivity of our results to specific model setups, including GCE model choice, atomic diffusion mitigation, and the assumed baseline. The complete matrix of model setups is listed in Table~\ref{tab:test} in Appendix~\ref{app:test}. We adopt the [X/Fe] setup with the solar baseline as the fiducial (denoted as Sun-[X/Fe]) and base our main conclusions on the results from this fiducial setup. We justify this choice in the next subsection.

\subsection{Candidate selection and validation}
\label{validation_scheme}

We performed model comparisons using Bayes factors between the engulfment model and each of the null models. We did not adopt a fixed cutoff to determine whether the engulfment model is favored; instead, we performed simulations to identify a reasonable cutoff for each model pair with a procedure similar to \cite{behmard_planet_2023}. For each star and under each null model, we performed a parametric bootstrap to simulate 10 mock abundance vectors under the best-fitting null prediction with added Gaussian noise with variance $\sigma_X^2 = \sigma_{X,\rm obs}^2+\sigma_{\rm int}^2$. 

The 121 sets of abundance vectors (for 113 stars and 8 overlapping stars with two sets each) yielded 1210 mock realizations for each tested null model, resulting in 1210 log Bayes factors ($\Delta\ln Z(E-N)$ values). We adopted the largest log Bayes factor among these 1210 values and set it as the threshold for rejecting the null, corresponding to a false-positive probability of $\sim$1/1210. This mixes the Bedell and the PASTA sample, but is equivalent to adopting the largest log Bayes factor between the two samples as the threshold for both, which is a conservative choice.

We classified a star as an engulfment candidate if, when comparing with \textit{all} alternative nulls (e.g., for Sun-[X/Fe], they are $F$, $G_{\rm Bedell}$, and $G_{\rm GALAH}$), the log Bayes factor of either the bulk-Earth or CM-chondrite model exceeds its calibrated threshold (see more in Appendix~\ref{app:test}). The distributions of the mock log Bayes factors and the thresholds under the fiducial model setup (Sun-[X/Fe]) are shown in Figure~\ref{fig:mock_xfe}, and the results for all other model setups are provided in Appendix~\ref{mock_signals}.

We adopted the Sun-[X/Fe] model setup as the fiducial for two reasons. First, as described in Section~\ref{main_models}, [X/Fe] reduces the signatures of atomic diffusion (see Appendix~\ref{atomic_diffusion} for more details). Second, the Sun provides a physically motivated baseline null case for planet engulfment, since the present Solar System shows no evidence of recent planet-scale engulfment. Long-term dynamical integrations indicate that planet–Sun collisions are rare for the current terrestrial-planet architecture \citet{Laskar2009}, and the Sun is refractory-depleted rather than refractory-enhanced relative to many solar twins \citep[e.g.,][]{melendez_peculiar_2009}.

We further test this choice using a mirror experiment in which all abundance differences are ``mirrored", i.e., multiplied by $-1$ (on logarithmic scales). If the baseline star has no engulfment signatures (e.g., the Sun) and the fitted engulfed mass is constrained to be positive, then reversing the abundance vector should significantly reduce the Bayes factors of engulfment models versus the nulls, as engulfment only enhances abundances in stars by adding materials to the stellar envelope. Among all model setups with different baselines and abundance scales, only the Sun-[X/Fe] model setup shows the expected overall decrease in the log Bayes factors ($\Delta\ln Z(E-N)$). The full sets of mirror-test results are presented in Appendix~\ref{mirror}. This further justifies our choice to use Sun-[X/Fe] for candidate selection and to treat the remaining setups as robustness tests.

\begin{figure*}
    \centering
    \includegraphics[width = 0.95\linewidth]{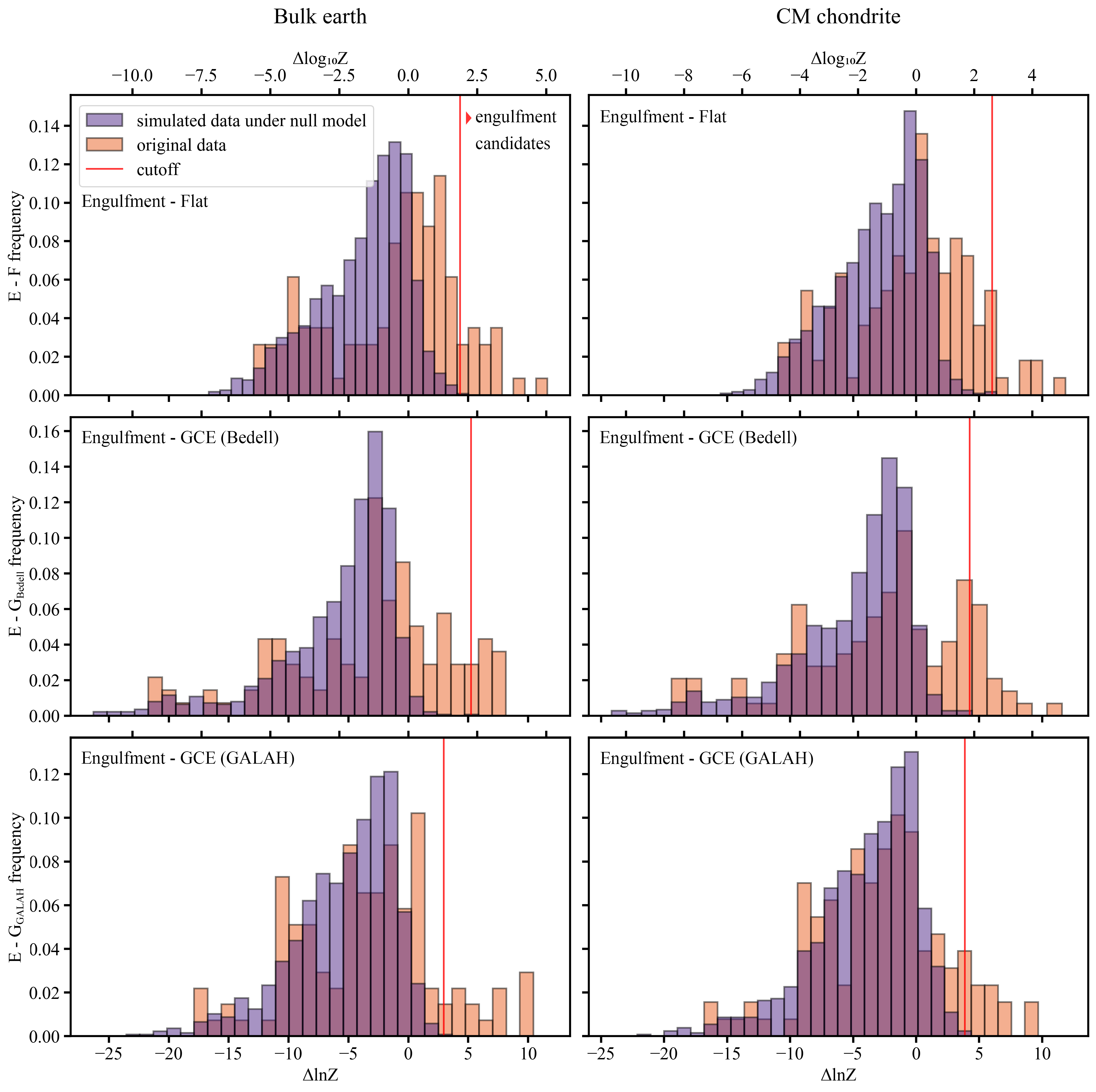}
    \caption{The distribution of \(\Delta \ln Z\) for the original data (orange) and the generated mock signals (purple) under the fiducial modeling setup (Sun-[X/Fe]). The two columns represent the two engulfed compositions (bulk earth on the left, and CM chondrite on the right), and the three rows represent the three models that are compared to the engulfment model. The red vertical line labels the cutoff value for each model comparison.}
    \label{fig:mock_xfe}
\end{figure*}

%%%%%%%%%%%%%%%%%%%%%%%%%%%%%%%%%%%%%%%%%%%%%%%%%%%%%%%%%%%%%%%%%%%%%%%%%%%%%%%%
\section{Results\label{result}}

% First, the main conclusion, with the main table and figure
\par
\begin{figure*}[h]
    \centering
        \includegraphics[width = 17cm]{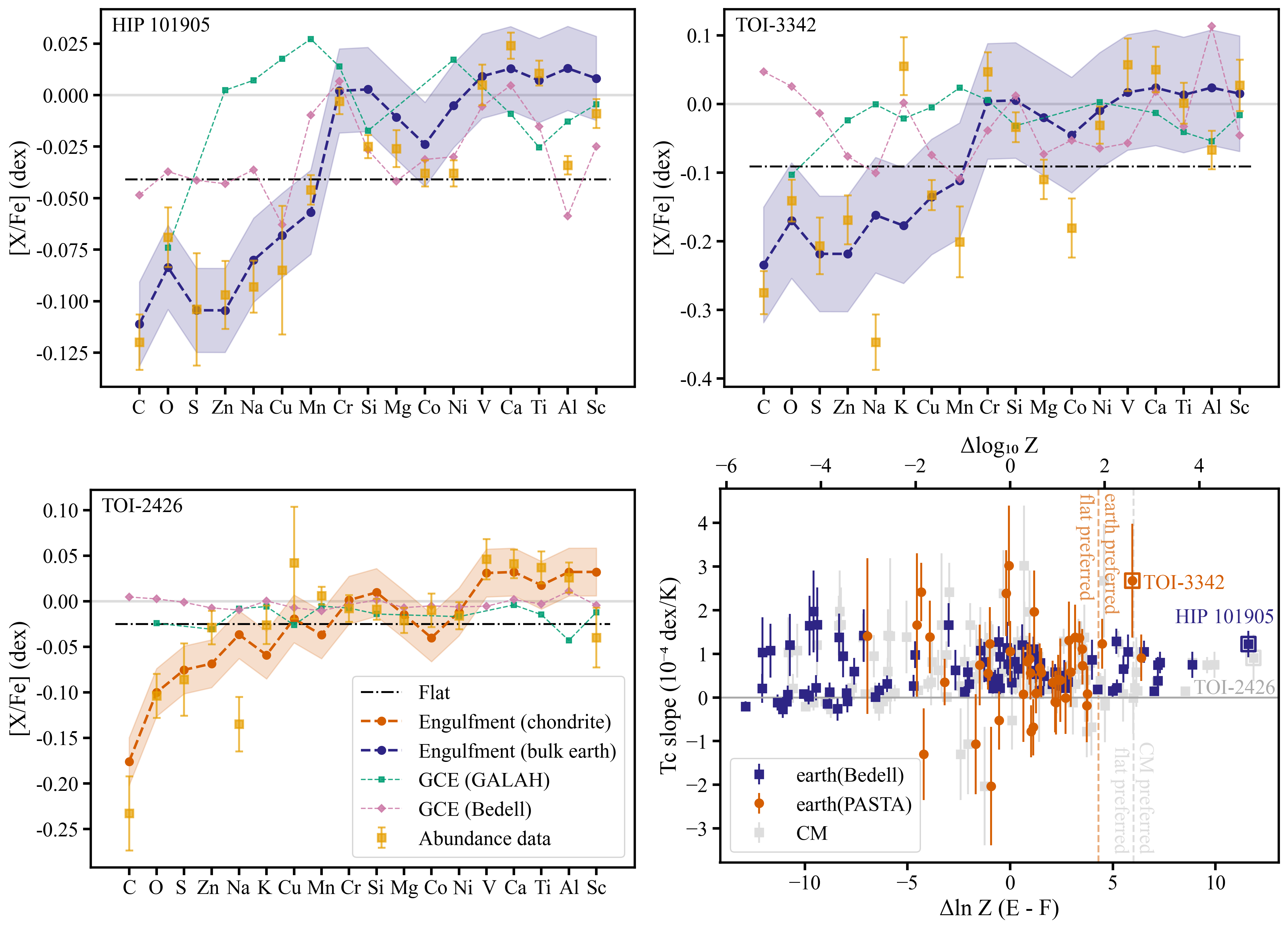}
        
    \caption{\label{fig:candidate}
    \textit{Top and lower left:} Abundance patterns ([X/Fe]) of the three candidates. The predicted abundance of the engulfment model is shown as the blue (bulk Earth) and red (CM chondrite) lines, with the shaded region demonstrating the \(1\sigma_{\text{int}}\) scatter. The best predicted abundance of the flat model, the GCE (GALAH) model, and the GCE (Bedell) model are shown as the black, green, and purple lines, respectively. \textit{Lower right:} \(T_{\text{c}}\) slope vs.~\(\Delta \ln Z (E - F)\) for the entire sample. The bulk-Earth points are red circles (PASTA) and blue squares (Bedell), and the cutoff value of \(\Delta \ln Z (E - F)\) is marked by the red vertical dashed line. The best engulfment candidates are highlighted using hollow squares and labeled with their names. The CM points and cutoff value are shown in gray. The error bars of \(\Delta \ln Z\) are not plotted as the errors (\(\sim 0.1\)) are too small to show. 
    }
\end{figure*}

Under our fiducial model setup (Sun-[X/Fe]) and model selection criteria, we identified three candidate stars with signatures of planet engulfment: TOI-3342, HIP 101905, and TOI-2426. These three stars also indicate some preference for the engulfment model in other modeling setups (green blocks in Table~\ref{tab:candidate}). HIP~101905 from the Bedell sample shows an excess of bulk Earth composition, and from the PASTA sample, TOI-3342 with bulk Earth and TOI-2426 with CM chondrite composition. The best-fit engulfment models for the three candidates, along with the other models, are shown in Figure \ref{fig:candidate}.

% Secondary conclusion with another plausible candidate
\par
In addition, HIP~30502 emerges under the mean-[X/Fe] model setup for both bulk Earth and CM chondrite compositions. Table \ref{tab:candidate} lists the $\Delta \ln Z$ values of all four candidates under all model setups, with green highlighting the values above the cutoff. 

% Further exploring the sample with loosened model selection criteria
\par
We identified 14 additional optimistic candidates by adopting a more permissive selection criterion and requiring the engulfment model to outperform only the flat model. We treat these as optimistic candidates because their patterns may be explained by GCE (see Appendix~\ref{app:lnz} for more details).

\begin{figure*}
     \centering
     \includegraphics[width=\linewidth]{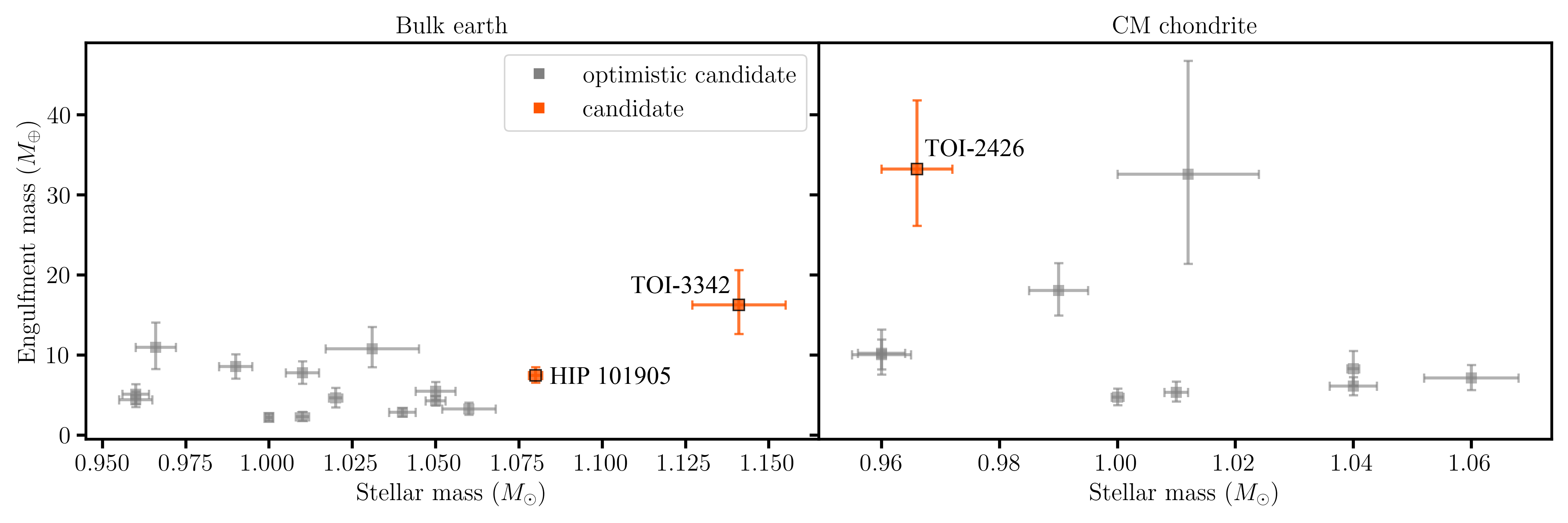}
     \caption{Engulfed mass vs. stellar mass under different engulfed compositions (left: bulk Earth, right: CM chondrite). The best candidates (orange) and optimistic candidates (gray) selected under the Sun-[X/Fe] model setup are shown in squares. }
     \label{fig:mass}
 \end{figure*}

% Estimates of engulfed mass
In Figure~\ref{fig:mass}, we plot the best-fit engulfed mass versus the stellar mass for the three best candidates (orange) and the other optimistic candidates (gray). We derived the engulfed mass using the best-fit equivalent engulfed mass (Equation~\ref{eqn:meq}) and an estimate of the convective zone mass of each star using a MESA grid by \citet{cheng_mesa_2026}. The three best candidates have engulfed masses of \(16\pm4\), \(7.5\pm1.0\), and \(33_{-7}^{+9}\) $M_\oplus$ for TOI-3342, HIP 101905, and TOI-2426, respectively.

% Does not look like systematics
Among all candidate stars, no best-fit engulfed mass exceeds 35 $M_\oplus$, which is well below the rough estimate of total solids in the minimum-mass solar nebula (e.g., \citealt{Hayashi1985}), although a larger engulfed mass would be easy to fit and would be favored with a large Bayesian evidence. Figure~\ref{fig:mass} also shows no correlation between stellar mass and the best-fit engulfed mass, which is physically reasonable.%, whereas abundance patterns caused by stellar systematics may exhibit such a correlation. %although more massive stars have significantly less massive convective zones than lower mass stars (even within our sample, which does not have a large span of stellar mass, the mass range of the convection zones is roughly \(0.01\)--\(0.03 M_\odot\); see Cheng et al. 2026 for more). 

% Engulfment rate
\par
In summary, three stars in our sample of 113 stars meet the criteria for planet engulfment candidates, indicating a raw detection rate of engulfment signatures among (non-co-natal) solar twins and analogs of $\sim$2.7\%. This is a raw estimate without sensitivity or false-positive corrections, but it is an overall representative number to quote; we present a more careful calculation in Appendix~\ref{app:rate}. We discuss further the detection rate of planet engulfment signatures in the next section.
%This estimate also does not take into account the fact that the Bedell sample and the PASTA sample need to be treated as two independent samples, as their abundance measurements have significantly different precision, with potential systematic offsets among some elements as well. Counting separately, there are 2 engulfment candidates out of 40 planet-host stars in PASTA (42 stars in total) and 1 out of 68 stars with no known planets in the Bedell sample (79 stars in total). 

\begin{deluxetable}{c|cccc|ccc|ccc|cc}
\tablecaption{List of Best Engulfment Candidate Stars and Their \(\Delta \ln Z\) Values}

\tablehead{
\multicolumn{1}{c}{} & \multicolumn{7}{c}{\(\Delta \ln Z\) for Baseline: the Sun} & \multicolumn{5}{c}{\(\Delta \ln Z\) for Baseline: sample mean} \\
\multicolumn{1}{c}{} & \multicolumn{4}{c}{[X/H]} & \multicolumn{3}{c}{\textbf{[X/Fe]}} & \multicolumn{3}{c}{[X/H]} & \multicolumn{2}{c}{[X/Fe]} \\
\colhead{Name} & \colhead{\(\text{E}\! - \!\text{F}\)} & \colhead{\(\text{E}^\prime\! - \!\text{Gb}\)} & \colhead{\(\text{E}^{\prime\prime}\! - \!\text{Gg}\)} & \colhead{\(\text{Eo}\! - \!\text{F}\)} & \colhead{\(\text{E}\! - \!\text{F}\)} & \colhead{\(\text{E}\! - \!\text{Gb}\)} & \colhead{\(\text{E}^{\prime\prime}\! - \!\text{Gg}\)} & \colhead{\(\text{E}\! - \!\text{F}\)} & \colhead{\(\text{E}^\prime\! - \!\text{Gb}\)} & \colhead{\(\text{Eo}\! - \!\text{F}\)} & \colhead{\(\text{E}\! - \!\text{F}\)} & \colhead{\(\text{E}\! - \!\text{Gb}\)}
}
\startdata
\multicolumn{1}{c|}{\textbf{cutoff/earth}} & \textbf{5.3} & \textbf{1.6} & \textbf{4.4} & \textbf{3.0} & \textbf{4.3} & \textbf{5.3} & {\textbf{3.0}} & {\textbf{3.3}} & {\textbf{1.8}} & {\textbf{4.0}} & {\textbf{4.2}} & {\textbf{1.8}} \\
\hline
    TOI-3342 & 3.9   & \cellcolor[rgb]{ .573,  .816,  .314}4.4 & 3.9   & \cellcolor[rgb]{ .573,  .816,  .314}4.3 & \cellcolor[rgb]{ .573,  .816,  .314}6.0 & \cellcolor[rgb]{ .573,  .816,  .314}6.9 & \cellcolor[rgb]{ .573,  .816,  .314}4.2 & 2.4   & -0.5  & 1.8   & \cellcolor[rgb]{ .573,  .816,  .314}4.4 & 1.7 \\
    HIP101905 & \cellcolor[rgb]{ .573,  .816,  .314}13.3 & \cellcolor[rgb]{ .573,  .816,  .314}7.4 & \cellcolor[rgb]{ .573,  .816,  .314}11.0 & \cellcolor[rgb]{ .573,  .816,  .314}11.4 & \cellcolor[rgb]{ .573,  .816,  .314}11.6 & \cellcolor[rgb]{ .573,  .816,  .314}6.6 & \cellcolor[rgb]{ .573,  .816,  .314}9.3 & \cellcolor[rgb]{ .573,  .816,  .314}9.6 & \cellcolor[rgb]{ .573,  .816,  .314}4.0 & \cellcolor[rgb]{ .573,  .816,  .314}6.7 & \cellcolor[rgb]{ .573,  .816,  .314}4.8 & 0.3 \\
    (HIP30502)  & -17.7 & -19.3 & -12.7 & \cellcolor[rgb]{ .573,  .816,  .314}4.8 & \cellcolor[rgb]{ .573,  .816,  .314}7.2 & 4.6   & 0.6   & -26.1 & -27.0 & \cellcolor[rgb]{ .573,  .816,  .314}4.4 & \cellcolor[rgb]{ .573,  .816,  .314}7.9 & \cellcolor[rgb]{ .573,  .816,  .314}5.1 \\ \hline
    \multicolumn{1}{c|}{\textbf{cutoff/CM}} & {\textbf{4.2}} & {\textbf{3.5}} & {\textbf{7.6}} & {\textbf{2.9}} & {\textbf{6.0}} & {\textbf{4.2}} & {\textbf{3.9}} & {\textbf{4.5}} & {\textbf{3.6}} & {\textbf{3.0}} & {\textbf{5.4}} & {\textbf{2.4}} \\ \hline
    TOI-2426 & -6.5  & -6.1  & -2.8  & \cellcolor[rgb]{ .573,  .816,  .314}8.8 & \cellcolor[rgb]{ .573,  .816,  .314}11.8 & \cellcolor[rgb]{ .573,  .816,  .314}11.6 & \cellcolor[rgb]{ .573,  .816,  .314}5.0 & -13.7 & -14.0 & 3.2   & 3.9   & \cellcolor[rgb]{ .573,  .816,  .314}2.5 \\
    (HIP30502)  & -17.6 & -19.2 & -12.9 & \cellcolor[rgb]{ .573,  .816,  .314}6.2 & \cellcolor[rgb]{ .573,  .816,  .314}9.6 & \cellcolor[rgb]{ .573,  .816,  .314}7.0 & 2.1   & -25.9 & -26.9 & \cellcolor[rgb]{ .573,  .816,  .314}7.3 & \cellcolor[rgb]{ .573,  .816,  .314}8.4 & \cellcolor[rgb]{ .573,  .816,  .314}5.6 \\
\enddata
\label{tab:candidate}
\tablecomments{
The table lists the stars that have passed all the tests under at least one model setup (i.e., our three best candidates under the fiducial Sun-[X/Fe] model setup, and HIP~30502 passing under the mean-[X/Fe] setup). The log Bayesian evidence differences or log Bayes factors \(\Delta \ln Z\) are listed for each star. The column head \(\text{A} - \text{B}\) refers to \(\Delta \ln Z(\text{Model A} - \text{Model B})\). The abbreviations Gb, Gg, and Eo represent \(G_{\rm Bedell}\), \(G_{\rm GALAH}\), and \(E_{\rm offset}\), respectively. The prime of the letter E (E$^\prime$ or E$^{\prime\prime}$) means that the engulfment model fitting is performed with a subset of elements to match the same set of elements being fitted in the null model (i.e., \(G_{\rm Bedell}\), \(G_{\rm GALAH}\); see Section~\ref{main_models} for details). The rows of cutoff values under each composition (``earth'' for the bulk earth and ``CM'' for the CM chondrite) are followed by their candidates, which at least have all cells green (i.e., passing the cutoff) under at least one model setup. In this table, the star IDs from the Bedell sample all begin with ``HIP'', whereas those from PASTA begin with ``TOI''. See Section~\ref{result} for more details.
}

\end{deluxetable}

%%%%%%%%%%%%%%%%%%%%%%%%%%%%%%%%%%%%%%%%%%%%%%%%%%%%%%%%%%%%%%%%%%%%%%%%%%%%%%%%
\section{Discussion and Conclusion\label{discussion}}

\subsection{Detection Rate of Planet Engulfment Signatures}\label{diss:rate}

% raw rate estimate
Under our fiducial Sun-[X/Fe] model setup, three of the 113 unique stars satisfy all selection criteria and thus represent our best engulfment candidates. This yields a raw estimate for the planet-engulfment detection rate of 2.7\% under our methodology, assuming that planet hosts and non-planet hosts share the same rate and ignoring other systematic differences between the Bedell and PASTA samples. Within the Bedell sample alone (a more generic nearby solar twin/analog sample), the raw detection rate is 1/79, or 1.3\%. We note that these ($\sim$1--3\%) estimates of \textit{detection} rate of planet engulfment signatures may represent conservative estimates for the \textit{occurrence} rate of engulfment events, since the three candidates exhibit planet engulfment signatures as the \textit{predominant} feature in their abundance patterns, while other stars with bona fide planet engulfment events could have their spectral signatures diluted or faded away already (e.g., due to small engulfed masses or mixture with GCE or random factors). Appendix~\ref{app:rate} presents a more detailed calculation and discussion on the detection rate calculation and explains why we choose to present the raw estimate of 1--3\% as a representative quote.

% compared with Behmard and Liu
Our estimated detection rate of $\sim$1--3\% is broadly consistent with \cite{behmard_planet_2023}, who reported an upper limit of 4.9\%, as well as \cite{liu_at_2024} with an estimate of one in a dozen ($\sim$8\%). The comparison is not one-to-one: previous studies on binaries or co-natal stars benefit from having the companion star as a control, whereas our non-co-natal sample requires explicit modeling of GCE and an overall metallicity offset. 

% Optimistic rate compared with Spina, with a word of caution
A more permissive selection, requiring only that the engulfment model outperform the flat model, yields 17 total candidates, or $\sim$15\% of the sample, closer to the $\sim$20--35\% engulfment probability inferred by \citet{spina_chemical_2021} from chemical anomalies in binary systems. The difference between our conservative and optimistic rates underscores the importance of the null model: candidate fractions based solely on deviations from a flat abundance pattern can be substantially higher than those that require the engulfment model to outperform GCE alternatives.

% Planet hosts have more engulfment events?
Finally, the present sample also hints, but does not establish, that engulfment signatures may be more commonly detected among planet hosts. Two of the three fiducial candidates are from the PASTA planet-host sample, whereas only one is from the Bedell sample and is not known to host any planets. The raw fractions are therefore 2/40 among PASTA planet-host stars and 1/68 among Bedell stars with no known planets. However, these numbers should not be over-interpreted because the two samples have different selection functions, abundance precision, planet-detection completeness, and possible abundance systematics. A robust comparison between planet hosts and non-hosts will require a larger, homogeneous sample with uniform abundance analysis and well-characterized completeness of planet detection (see more in Appendix \ref{app:rate}).

%%%%%%%%%%%%%%%%%%%%%%%%
\subsection{Going Beyond the $T_{\text{c}}$ Slope \label{diss:tc}}

\begin{figure*}
    \centering
    \includegraphics[width=8cm]{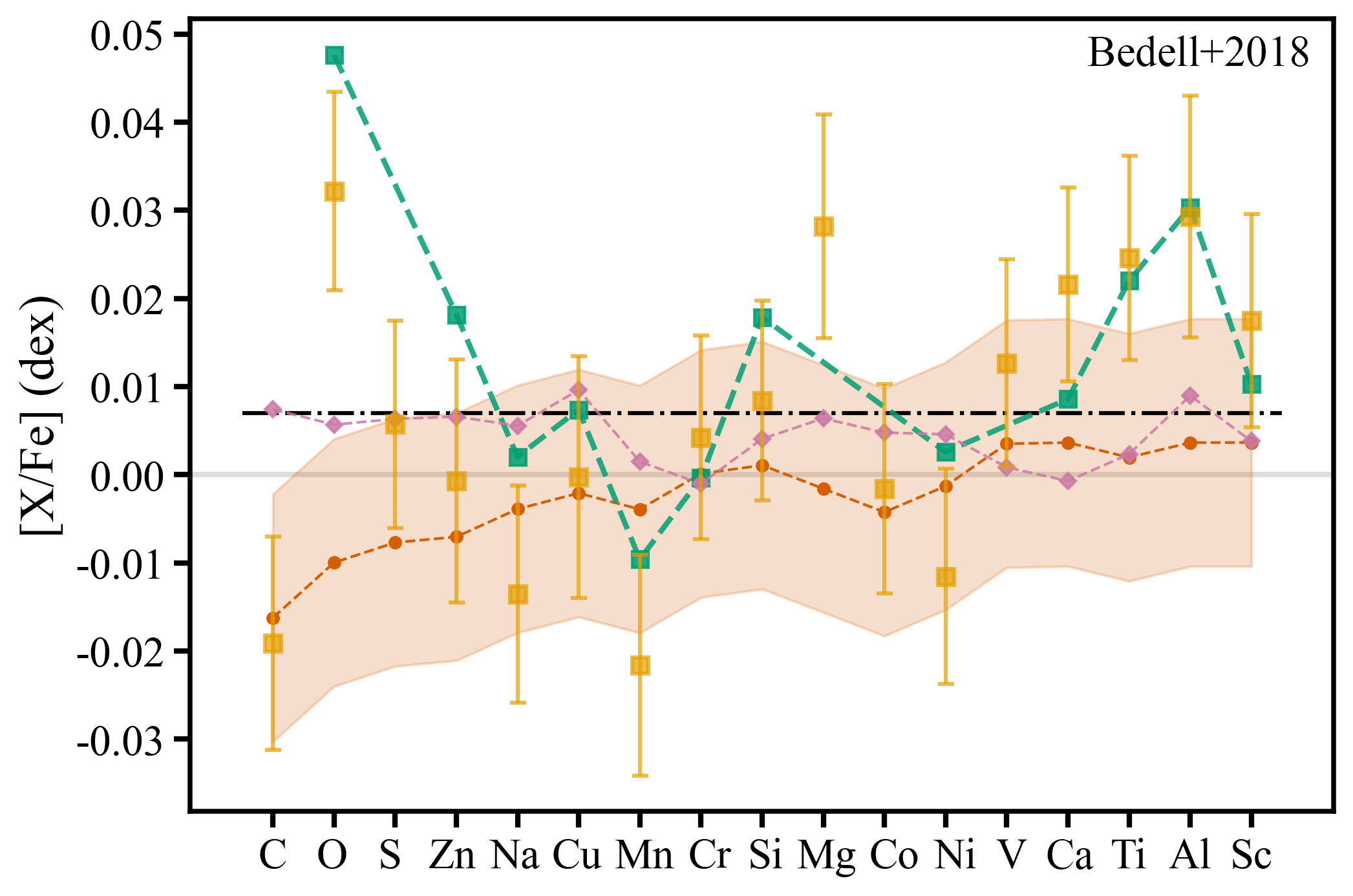}
    \includegraphics[width=8cm]{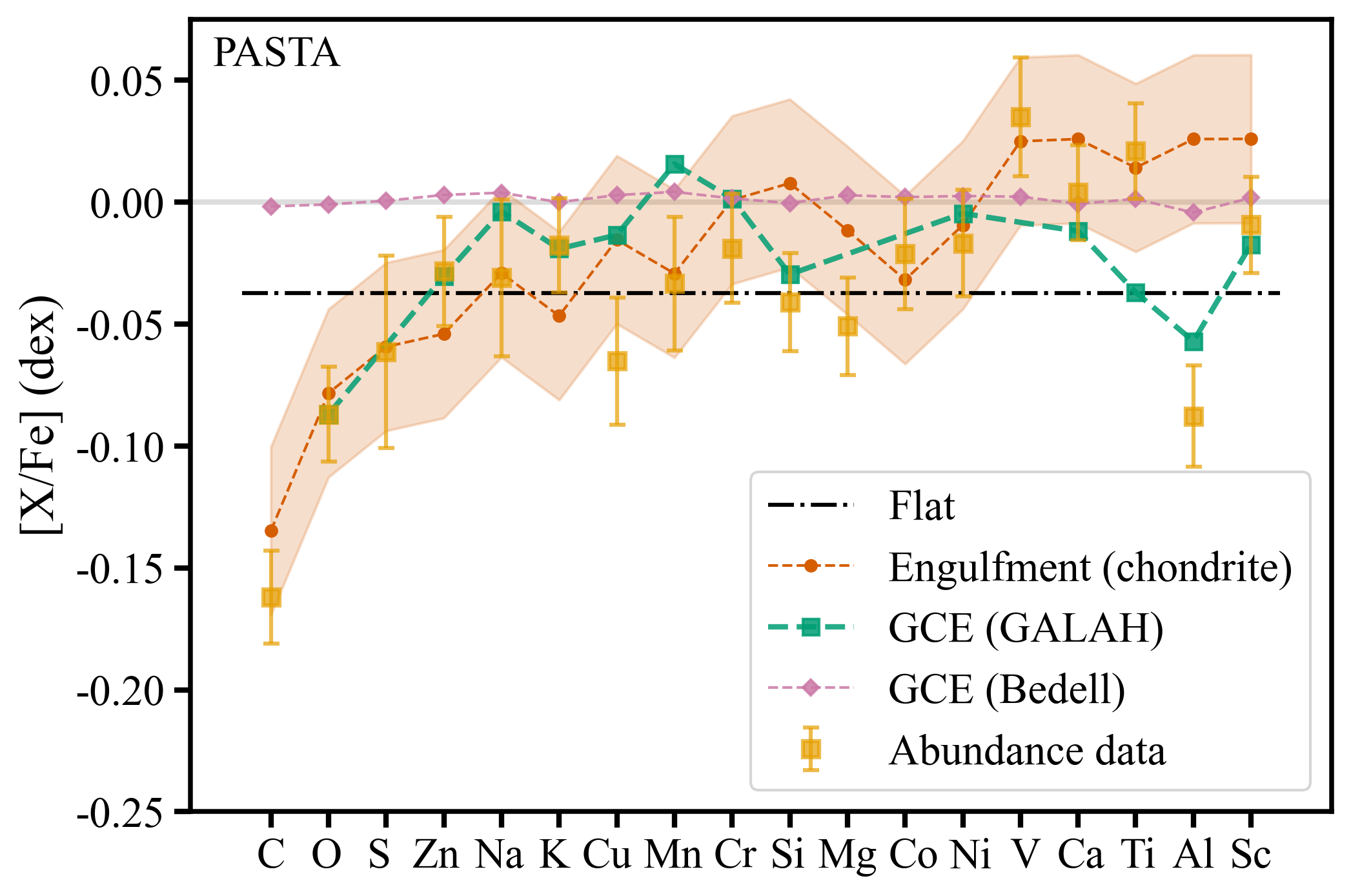}
    \caption{Abundance differences ([X/Fe]) of the two sample means. Same as Figure \ref{fig:candidate}. Notice that the abundance scales are different.}
    \label{fig:sample_mean}
\end{figure*}

% high Tc slope != engulfment, Figure 2 lower right
Trends between abundance and condensation temperature, commonly summarized by a $T_{\rm c}$ slope, have been widely used to describe the refractory depletion of the Sun relative to solar twins and to search for possible planet-formation signatures \citep{melendez_peculiar_2009,bedell_abundance_2014}. However, a $T_{\rm c}$ slope reduces a multi-element abundance pattern to a single number. This compression can obscure the detailed element-by-element structure expected from engulfment, GCE, or other processes. As shown in the lower-right panel of Figure~\ref{fig:candidate}, stars with large $\Delta \ln Z(\text{E}- \text{F})$ are not simply those with the steepest $T_{\rm c}$ slopes. In other words, a large or small $T_{\rm c}$ slope is neither a necessary nor sufficient condition for an abundance pattern to match an engulfment template.

% sample average
The same limitation appears in the sample-mean abundance patterns shown in Figure~\ref{fig:sample_mean}. For a fair comparison in Figure~\ref{fig:sample_mean}, we exclude the 11 potential and confirmed planet hosts (including the brown dwarf hosts like HIP~5301) from the Bedell sample. For the PASTA sample, we exclude the 2 non-planet hosts and another 9 stars whose \(T_{\rm eff}\) and [Fe/H] are far beyond the range of the Bedell sample, so that the [X/H] range is \(\pm 0.15 \mathrm{dex}\) and the \(T_{\text{eff}}\) range shrinks to \(5630\mathrm K\sim 6000\mathrm{K}\). 

The two average abundance patterns both show a refractory depletion \(T_c\) trend, but the PASTA one exhibits greater abundance fluctuations. Interestingly, the PASTA sample mean shows a marginal chondrite engulfment signature with a \(\Delta \ln Z \) of \(4.7\). However, the GALAH GCE model fits the patterns even better for both sample means. This pattern, whether caused by engulfment or GCE, would not be discovered through the \(T_c\) slope. In addition, the condensation sequence, and thus the \(T_c\) values, for the same set of elements could be affected by systematic differences in stellar abundances \citep{Spaargaren2025, Zaveri2026}, which may therefore obscure the observed \(T_c\)-based patterns. 
Therefore, we suggest that $T_{\rm c}$ slopes should be treated as a useful descriptive summary statistic rather than as a standalone diagnostic for planet engulfment. Template-based or more detailed modeling approaches preserve more chemical information and can distinguish abundance patterns that would otherwise be collapsed into similar $T_{\rm c}$ slopes or volatile-refractory ratios.

\subsection{Limitations and Caveats}\label{diss:caveat}

%\par
This work focuses on explaining the differences in abundance patterns between solar analogs/twins and the Sun in the context of planet engulfment. However, other effects related to planet formation can also introduce such differences in abundances, such as the formation of planets \citep[e.g.,][]{chambers_slar_2010,kunitomo_revisiting_2018,booth_fingerprints_2020}. These effects can be degenerate in observational signatures, and one possible path forward to distinguish the exact origin is to establish a large sample of solar twins and analogs with high-precision abundance measurements, well-characterized planetary architectures, and, ideally, some inference of the planetary system's dynamical history.

\par
Another caveat of this study is that model selection based on Bayes factors and Bayesian evidence can be prior-sensitive. A wider prior distribution may also dilute significance. We used an uninformative uniform prior over the engulfed mass, with reasonable but somewhat arbitrary bounds. However, if we change the upper limit of the prior distribution from \(80M_\oplus\) to \(50M_\oplus\)(approximately \(80e^{-0.5}\)) or \(130M_\oplus\)(approximately \(80e^{0.5}\)), then \(\ln Z\) increases or decreases by about 0.5. Given the differences in variables across models and the fact that priors are often chosen to be arbitrarily uninformative, it is difficult to compare any two models on an absolutely fair baseline. This is why we chose to determine the cutoff $\Delta \ln Z$ values for different model comparisons through the mock signals, which provide a meaningful baseline and a symbolic p-value for each test.

%%%%%%%%%%%%%%%%%%%%%%%%
\subsection{Conclusion and Future Work}

\par
In summary, we found one Earth-like engulfment candidate in both the Bedell sample and the PASTA sample, and one CM chondrite-like engulfment candidate in the PASTA sample, which indicates a lower limit on the occurrence of planet engulfment signatures of $\sim$1--3\%. The candidates from PASTA, TOI-2426, and TOI-3342 each have a high-probability planet candidate. It is worthwhile to conduct further observations to confirm them (and the planet candidates around the optimistic candidates) and to identify additional key characteristics that could support such a violent event.

\par
Additionally, further investigation into alternative effects, such as GCE and atomic diffusion, is required, as the available elements and predictive ability cover only a subset of the observed elements to date. We also need a larger set of homogeneous, high-resolution spectra with uniform analyses for solar twins and analogs to impose a more precise constraint on the engulfment rate.

%% Please use the acknowledgment and contribution environments. This will 
%% be anonymized when the "anonymous" style option is used. 
\begin{acknowledgments}

This work is supported by the National Key R\&D Program of China under Grant No.~2025YFE0102100. ZC, SXW, and ZG acknowledge support from the China-Chile Joint Research Fund (CCJRF No.2301) and the Chinese Academy of Sciences South America Center for Astronomy (CASSACA) Key Research Project E52H540301. This work is also supported by the National Key R\&D Program of China under Grant No. 2024YFA1611801, the Science and Technology Commission of Shanghai Municipality under Grant No. 25ZR1402244, and the Shanghai Jiao Tong University Funds Program No. AF4260012. HSW acknowledges support from the Carlsberg Foundation through the FIRSTATMO project. ZG is supported by FONDECYT Iniciacion project 11260176. MB is supported by the Australian Research Council (ARC) Center of Excellence for Gravitational Wave Discovery (OzGrav) through project number CE230100016, and the Commonwealth through an Australian Government Research Training Program Scholarship [DOI: https://doi.org/10.82133/C42F-K220]. JO acknowledges support from ANID Becas/Doctorado nacional/2026-21262687. YST acknowledges support from the National Science Foundation under Grant No. AST-2406729 and a Humboldt Research Award from the Alexander von Humboldt Foundation.

The authors used ChatGPT (OpenAI, GPT-5.5 Thinking) for language polishing and editorial suggestions. The authors reviewed and edited all AI-assisted text and take full responsibility for the content of this manuscript.

\end{acknowledgments}

%\begin{contribution}
%%This section gives authors the space to recognize author contributions. The text inside this environment is NOT counted towards the total word quanta. At a minimum, manuscripts are expected to include this text:

% need content here......

%% But authors are expected to provide more specific details, e.g. 
%%
%%SC was responsible for writing and submitting the manuscript.
%%WWM came up with the initial research concept and edited the manuscript.
%%OTS obtained the funding and edited the manuscript.
%%EBF provided the formal analysis and validation. He also edited the manuscript.
%%GEH Supervised the undergraduates, wrote the software and administers the project github and Zenodo repositories.
%%
%% Authors can use the Contributor Role Taxonomy (CRediT) at
%% for ideas on how write a good statement tailored to their needs.

%\end{contribution}


\facilities{ESO:3.6m (HARPS), {\it Magellan}:Clay (MIKE).}

\software{astropy \citep{2013A&A...558A..33A, collaboration_astropy_2022}, 
dynesty \citep{speagle_dynesty_2020}.
}


%%%%%%%%%%%%%%%%%%%%%%%%%%%%%%%%%%%%%%%%%%%%%%%%%%%%%%%%%%%%%%%%%%%%%%%%%%%%%%%%
\appendix

\section{Sample Construction and Catalog Details}\label{app:data}

\subsection{Source Samples and Spectroscopic Data}

The stellar parameters and abundances used in this work were compiled from \citet{bedell_chemical_2018}, \citet{spina_temporal_2018}, and \citet{sun_planets_2025,sun_planets_2025-1}. The observational characteristics and parameter coverage of the two samples differ and are summarized below. The data used in this work are provided in Table \ref{tab:star_data}.

\par
The spectral data for the Bedell sample of 79 stars were taken with the High Accuracy Radial velocity Planet Searcher (HARPS; $R=$115,000) and the Magellan II/MIKE spectrograph
($R=$83,000--65,000 for red or blue band). The spectra from HARPS have, on average, an S/N of approximately \(800~\mathrm{pix}^{-1}\) at \(600~ \mathrm{nm}\), and those from MIKE have \(400~\mathrm{pix}^{-1}\) at \(600~ \mathrm{nm}\). The spectra from MIKE were only used for 11 lines of C, CH, and O. The range of stellar parameters compared with the Sun is typically: \(\pm 150 ~\mathrm{K}\) for \(T_{\text{eff}}\); between \(4.1\) dex and \(4.55 ~\mathrm{dex}\) for \(\log g\); and \(\pm 0.15 ~\mathrm{dex}\) for [Fe/H].

\par
The PASTA sample of 42 stars, all observed with Magellan II/MIKE, has a typical S/N of \(200~\mathrm{pix}^{-1}\) at \(600~ \mathrm{nm}\). Compared with the Sun, the range of \(T_{\text{eff}}\) is roughly \(\pm 250 ~\mathrm{K}\); the range of \(\log g\) is roughly \(\pm 0.3 ~\mathrm{dex}\); and the range of [Fe/H] is roughly \(\pm 0.4 ~\mathrm{dex}\).

\par
The eight overlapping stars between the two samples are: TOI-1055 (HIP 96160), HD 75302 (HIP 43297), HIP 44713, HIP 54287, HD 20782 (HIP 15527), HD 88072 (HIP 49756), HD 42618 (HIP 29432), and HIP 25670, which are labeled as 1 in the ``overlap" column in Table~\ref{tab:star_data}. The two non-planet-hosting stars in the PASTA sample, which were taken for calibration and comparison purposes, are HIP 44713 and HIP 54287.

\subsection{Abundance Compilation}

\par 
The abundance data from the Bedell sample included different ions/molecules for the elements Sc, Ti, Cr, and C, which we merged using weighted averages (first converting numbers of atoms for averaging, then converting back to [X/H]), incorporating the reported uncertainties, to derive the total abundance of each atomic species.

\par
We excluded the abundance data for elements with atomic numbers greater than 30 in the samples because their abundances were derived from synthesis rather than through equivalent width (EW) measurements. The available elements are C, O, Na, Mg, Al, Si, S, K, Ca, Sc, Ti, V, Cr, Mn, Fe, Co, Ni, Cu, and Zn, except that the Bedell sample does not include K.

\subsection{Planet and Companion Classification}

\par
For each star, we compiled confirmed planets from the NASA Exoplanet Archive and additional listed companions from SIMBAD. PASTA targets without confirmed planets were retained as planet hosts when their TOIs met the high-probability selection adopted by \citet{sun_planets_2025}. Objects classified as brown dwarf ($M>13$ Jupiter mass) hosts were included in the planet-host category for sample-demographic comparisons.

\par
The total number of known planets for each star is recorded in column 6 of Table~\ref{tab:star_data}. This number is taken as the maximum of the NASA Exoplanet Archive and SIMBAD records, including known brown dwarfs (e.g., HIP 5301b). In total, there are 45 stars with known planets or low-mass brown dwarfs: 11 from the Bedell sample and 40 from the PASTA sample, including 6 overlapping targets. 

The ``non-planet-host" stars in the sample do not represent a sample of stars without planets. Rather, they are a sample of stars without \textit{known} planets, which could host planets yet to be detected (e.g., beyond the current detection limit or not sufficiently surveyed) or no planet at all, and thus they represent a generic yet heterogeneous comparison sample to the known planet-host stars.

\startlongtable

\begin{deluxetable}{cccc}
\tablecaption{Catalog Parameters for Solar Analogs Used in this Work}
\tablehead{
   \colhead{Row number} & \colhead{Column} & \colhead{Units} & \colhead{Description}
}
\startdata
    1     & \multicolumn{1}{l}{gaia\_dr3\_id} &       & \multicolumn{1}{l}{Gaia DR3 Source ID} \\
    2     & \multicolumn{1}{l}{source} &       & \multicolumn{1}{l}{Survey source} \\
    3     & \multicolumn{1}{l}{id} &       & \multicolumn{1}{l}{Star identifier used in source work} \\
    4     & \multicolumn{1}{l}{overlap} &       & \multicolumn{1}{l}{Overlapping star flag (1 = Yes, 0 = No)} \\
    5     & \multicolumn{1}{l}{debris\_disk} &       & \multicolumn{1}{l}{Debris disk star flag (1 = Yes, 0 = No)} \\
    6     & \multicolumn{1}{l}{planet\tablenotemark{a}} &       & \multicolumn{1}{l}{Number of known planets} \\
    7     & \multicolumn{1}{l}{ra} & \multicolumn{1}{l}{degree} & \multicolumn{1}{l}{Right ascension} \\
    8     & \multicolumn{1}{l}{dec} & \multicolumn{1}{l}{degree} & \multicolumn{1}{l}{Declination} \\
    9    & \multicolumn{1}{l}{teff} & \multicolumn{1}{l}{K} & \multicolumn{1}{l}{Effective temperature} \\
    10    & \multicolumn{1}{l}{e\_teff} & \multicolumn{1}{l}{K} & \multicolumn{1}{l}{Uncertainty in effective temperature} \\
    11    & \multicolumn{1}{l}{logg} & \multicolumn{1}{l}{dex} & \multicolumn{1}{l}{Surface gravity} \\
    12    & \multicolumn{1}{l}{e\_logg} & \multicolumn{1}{l}{dex} & \multicolumn{1}{l}{Uncertainty in surface gravity} \\
    13    & \multicolumn{1}{l}{mass} & \multicolumn{1}{l}{Msun} & \multicolumn{1}{l}{Stellar mass} \\
    14    & \multicolumn{1}{l}{e\_mass} & \multicolumn{1}{l}{Msun} & \multicolumn{1}{l}{Uncertainty in stellar mass} \\
    15    & \multicolumn{1}{l}{age} & \multicolumn{1}{l}{Gyr} & \multicolumn{1}{l}{Stellar age} \\
    16    & \multicolumn{1}{l}{e\_age} & \multicolumn{1}{l}{Gyr} & \multicolumn{1}{l}{Uncertainty in stellar age} \\
    17    & \multicolumn{1}{l}{Fe\_h} & \multicolumn{1}{l}{dex} & \multicolumn{1}{l}{Iron abundance [Fe/H]} \\
    18    & \multicolumn{1}{l}{e\_Fe\_h} & \multicolumn{1}{l}{dex} & \multicolumn{1}{l}{Uncertainty in [Fe/H]} \\
    19    & \multicolumn{1}{l}{C\_h} & \multicolumn{1}{l}{dex} & \multicolumn{1}{l}{Carbon abundance [C/H]} \\
    20    & \multicolumn{1}{l}{e\_C\_h} & \multicolumn{1}{l}{dex} & \multicolumn{1}{l}{Uncertainty in [C/H]} \\
    21    & \multicolumn{1}{l}{O\_h} & \multicolumn{1}{l}{dex} & \multicolumn{1}{l}{Oxygen abundance [O/H]} \\
    22    & \multicolumn{1}{l}{e\_O\_h} & \multicolumn{1}{l}{dex} & \multicolumn{1}{l}{Uncertainty in [O/H]} \\
    23    & \multicolumn{1}{l}{Na\_h} & \multicolumn{1}{l}{dex} & \multicolumn{1}{l}{Sodium abundance [Na/H]} \\
    24    & \multicolumn{1}{l}{e\_Na\_h} & \multicolumn{1}{l}{dex} & \multicolumn{1}{l}{Uncertainty in [Na/H]} \\
    25    & \multicolumn{1}{l}{Mg\_h} & \multicolumn{1}{l}{dex} & \multicolumn{1}{l}{Magnesium abundance [Mg/H]} \\
    26    & \multicolumn{1}{l}{e\_Mg\_h} & \multicolumn{1}{l}{dex} & \multicolumn{1}{l}{Uncertainty in [Mg/H]} \\
    27    & \multicolumn{1}{l}{Al\_h} & \multicolumn{1}{l}{dex} & \multicolumn{1}{l}{Aluminum abundance [Al/H]} \\
    28    & \multicolumn{1}{l}{e\_Al\_h} & \multicolumn{1}{l}{dex} & \multicolumn{1}{l}{Uncertainty in [Al/H]} \\
    29    & \multicolumn{1}{l}{Si\_h} & \multicolumn{1}{l}{dex} & \multicolumn{1}{l}{Silicon abundance [Si/H]} \\
    30    & \multicolumn{1}{l}{e\_Si\_h} & \multicolumn{1}{l}{dex} & \multicolumn{1}{l}{Uncertainty in [Si/H]} \\
    31    & \multicolumn{1}{l}{S\_h} & \multicolumn{1}{l}{dex} & \multicolumn{1}{l}{Sulfer abundance [S/H]} \\
    32    & \multicolumn{1}{l}{e\_S\_h} & \multicolumn{1}{l}{dex} & \multicolumn{1}{l}{Uncertainty in [S/H]} \\
    33    & \multicolumn{1}{l}{K\_h} & \multicolumn{1}{l}{dex} & \multicolumn{1}{l}{Potassium abundance [K/H]} \\
    34    & \multicolumn{1}{l}{e\_K\_h} & \multicolumn{1}{l}{dex} & \multicolumn{1}{l}{Uncertainty in [K/H]} \\
    35    & \multicolumn{1}{l}{Ca\_h} & \multicolumn{1}{l}{dex} & \multicolumn{1}{l}{Calcium abundance [Ca/H]} \\
    36    & \multicolumn{1}{l}{e\_Ca\_h} & \multicolumn{1}{l}{dex} & \multicolumn{1}{l}{Uncertainty in [Ca/H]} \\
    37    & \multicolumn{1}{l}{Sc\_h} & \multicolumn{1}{l}{dex} & \multicolumn{1}{l}{Scandium abundance [Sc/H]} \\
    38    & \multicolumn{1}{l}{e\_Sc\_h} & \multicolumn{1}{l}{dex} & \multicolumn{1}{l}{Uncertainty in [Sc/H]} \\
    39    & \multicolumn{1}{l}{Ti\_h} & \multicolumn{1}{l}{dex} & \multicolumn{1}{l}{Titanium abundance [Ti/H]} \\
    40    & \multicolumn{1}{l}{e\_Ti\_h} & \multicolumn{1}{l}{dex} & \multicolumn{1}{l}{Uncertainty in [Ti/H]} \\
    41    & \multicolumn{1}{l}{V\_h} & \multicolumn{1}{l}{dex} & \multicolumn{1}{l}{Vanadium abundance [V/H]} \\
    42    & \multicolumn{1}{l}{e\_V\_h} & \multicolumn{1}{l}{dex} & \multicolumn{1}{l}{Uncertainty in [V/H]} \\
    43    & \multicolumn{1}{l}{Cr\_h} & \multicolumn{1}{l}{dex} & \multicolumn{1}{l}{Chromium abundance [Cr/H]} \\
    44    & \multicolumn{1}{l}{e\_Cr\_h} & \multicolumn{1}{l}{dex} & \multicolumn{1}{l}{Uncertainty in [Cr/H]} \\
    45    & \multicolumn{1}{l}{Mn\_h} & \multicolumn{1}{l}{dex} & \multicolumn{1}{l}{Manganese abundance [Mn/H]} \\
    46    & \multicolumn{1}{l}{e\_Mn\_h} & \multicolumn{1}{l}{dex} & \multicolumn{1}{l}{Uncertainty in [Mn/H]} \\
    47    & \multicolumn{1}{l}{Co\_h} & \multicolumn{1}{l}{dex} & \multicolumn{1}{l}{Cobalt abundance [Co/H]} \\
    48    & \multicolumn{1}{l}{e\_Co\_h} & \multicolumn{1}{l}{dex} & \multicolumn{1}{l}{Uncertainty in [Co/H]} \\
    49    & \multicolumn{1}{l}{Ni\_h} & \multicolumn{1}{l}{dex} & \multicolumn{1}{l}{Nickel abundance [Ni/H]} \\
    50    & \multicolumn{1}{l}{e\_Ni\_h} & \multicolumn{1}{l}{dex} & \multicolumn{1}{l}{Uncertainty in [Ni/H]} \\
    51    & \multicolumn{1}{l}{Cu\_h} & \multicolumn{1}{l}{dex} & \multicolumn{1}{l}{Copper abundance [Cu/H]} \\
    52    & \multicolumn{1}{l}{e\_Cu\_h} & \multicolumn{1}{l}{dex} & \multicolumn{1}{l}{Uncertainty in [Cu/H]} \\
    53    & \multicolumn{1}{l}{Zn\_h} & \multicolumn{1}{l}{dex} & \multicolumn{1}{l}{Zinc abundance [Zn/H]} \\
    54    & \multicolumn{1}{l}{e\_Zn\_h} & \multicolumn{1}{l}{dex} & \multicolumn{1}{l}{Uncertainty in [Zn/H]} \\
\enddata
\label{tab:star_data}
\tablenotetext{a}{The number of planets is the maximum of SIMBAD ``child'' and the confirmed planets and TOIs from the NASA Exoplanet Archive. It includes non-confirmed planets and brown dwarfs.}
\tablecomments{Table \ref{tab:star_data} is published in its entirety in the electronic 
edition of the {\it Astrophysical Journal}.  A portion is shown here 
for guidance regarding its form and content.}
\end{deluxetable}

%%%%%%%%%%%%%%%%%%%%%%%%%%%%%%%%%%%%%%%%%%%%%%%%%%%%%%%%%%%%%%%%%%%%%%%%%%%%%%%%
\section{Detailed Model Information \label{detailed_model}}

This appendix provides the complete definitions of the abundance-pattern models, likelihood function, prior distributions, and model comparisons used in this work. The principal analysis is summarized in Section~\ref{method}.

\subsection{Modeling framework and notation}

For each star, we model a vector of elemental abundance differences relative to a reference abundance pattern. The reference is either the Sun or the mean abundance pattern of the corresponding survey. Our basic assumption is that, although many factors influence a star's chemical abundances, for some stars there may be a dominant factor that shapes the abundance pattern. These factors and thus the models used in this work include atomic diffusion, Galactic chemical evolution (GCE), and planet engulfment. To be specific:
\begin{equation}\label{eqn:frame}
    \Delta A_X = \varepsilon + \left\{
    \begin{array}{l}
        E(M_{\text{p}},\ldots)\text{, Engulfment;}\\
        G(\textit{age},\ldots)\text{, GCE;}\\
        D(\textit{age},\ldots)\text{, atomic diffusion.}\\
    \end{array}
    \right.
\end{equation}
Here \(\Delta A_X\) represents the abundance difference of element \(X\) relative to a certain baseline, for example, the Sun (i.e., then $\Delta A_X$ is simply [X/H]) or the respective sample mean (or any random star). \(E, G, \text{and } D\) are the specific models with their parameters (such as engulfed mass $M_{\rm p}$ or stellar age), and $\varepsilon$ represents residual abundance variation, which might be explained by intrinsic scatter between stars ($\sigma_{\rm int}$; see Equation \ref{eqn:likelyhood} in the next subsection). We note that the planet engulfment effect \(E\) is one of the possible planetary effects, and other possibilities are discussed in Section \ref{discussion}.

Our data do not support fitting all of these contributions simultaneously. Such a model would introduce many poorly constrained and partially degenerate parameters. We therefore test whether one model family provides the dominant description of the observed abundance pattern. This approximation does not imply that all other physical effects are absent. Rather, a candidate is selected when an engulfment model is favored over each of the non-engulfment alternatives considered.

\subsection{Likelihood and Bayesian evidence\label{app:bayesian_method}}

\par
For a given model that is described by the parameters \(\theta\) and the observed data \(D\), the Bayesian theorem provides the joint posterior distribution of the parameters:
\begin{equation}
    P(\theta\mid D) = \frac{P(D\mid \theta) \pi(\theta)}{P(D)}.
\end{equation}
The \(P(D\mid \theta)\) is the likelihood function \(L\) of the model; the \(\pi(\theta)\) is the prior distribution of the parameters; and the \(P(D) = \int_\theta P(D\mid \theta) \pi(\theta) \mathrm{d} \theta\) is the Bayesian evidence \(Z\).

\par
We used the same likelihood function \(L\) for all the models:
\begin{equation}\label{eqn:likelyhood}
\ln L = \mathit{norm} - \sum_j^N \frac{(\Delta A_{j,\text{observation}} - \Delta A_{j, \text{model}})^2}{2(\sigma_j^2 + \sigma^2_{\text{int}})}.
\end{equation}
\(\sigma_j\) is the measured uncertainty of the element \(j\), and \(\sigma_{\text{int}}\) is the intrinsic scatter with a uniform prior distribution on \((0, 1)\), which is introduced in all models to avoid overfitting and explain possible residuals. The \(N\) is the number of available elements; the logarithmic normalization term \(\textit{norm}\) is defined as:
\begin{equation}
\textit{norm} = -\frac{N\ln (2\pi)}{2} - \sum_j^N \ln\sqrt{\sigma_j^2 + \sigma^2_{\text{int}}}.
\end{equation}

\subsection{Abundance-pattern models\label{app:expression}}

\subsubsection{Flat model}\label{offset}

\par
The first model we consider is the scatter model, or the ``flat model'', with no correlation between abundances and atomic number or condensation temperature. \citet{behmard_planet_2023} and \citet{liu_at_2024} used this flat model (F) as the null hypothesis, i.e.,
\begin{equation}
    F_{\text{[X/H]}} = \mathit{offset}
\end{equation} 
This describes the intrinsic abundance difference between any two stars as dominated by an offset.

\subsubsection{Atomic diffusion\label{atomic_diffusion}}

\par
Beyond random scatter, even two stars born with identical abundances would evolve to have different patterns due to atomic diffusion. In this work, for [X/H] setups, we incorporate the effects of atomic diffusion into the flat model (i.e., replacing \(D\) with \(F\)). This choice is driven by the lack of available elements in existing atomic diffusion models (e.g., \citealt{dotter_influence_2017}) and the homogeneity of the stellar parameters in our samples.

\par
\begin{figure}
    \centering
    \includegraphics[width=\linewidth]{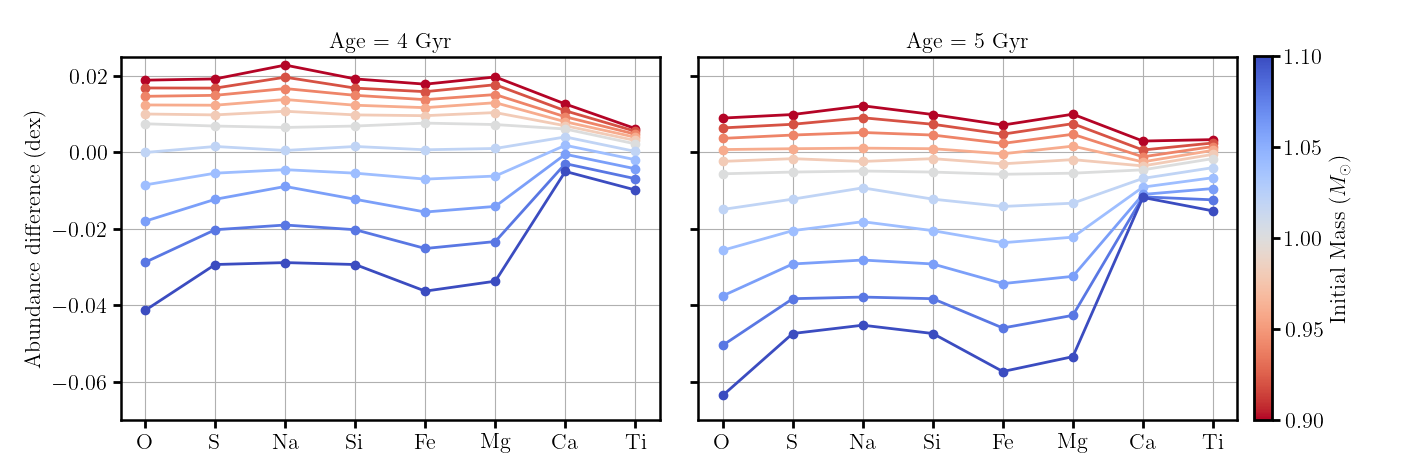}
    \caption{The abundance differences caused by atomic diffusion. The abundances are calculated based on the simulation results from \citet{dotter_influence_2017}. The baseline is at the age of 4.57 Gyr and the mass of \(1M_\odot\), where the model prediction is different from the actual solar abundance.}
    \label{fig:atomic}
\end{figure}

To verify this choice, we plot the effect of atomic diffusion, as calculated by stellar evolution models using MESA, in Figure \ref{fig:atomic} \citep{dotter_influence_2017}. Within the mass range of most stars in our samples, the abundance difference can indeed be approximated by a horizontal line (the flat model). However, due to the limited number of supported elements, the complete picture of the abundance difference remains unclear. 

When using [X/Fe] instead of [X/H], it is justified in previous works that the effect of atomic diffusion can be largely reduced \citep[e.g.,][]{dotter_influence_2017, moedas_atomic_2022}. In this case, the flat model describes stars that share a uniform value in [X/H] (plus scatter), while [Fe/H] is allowed to be much different from the other [X/H]. This is physically unreasonable. However, we still adopt the flat model as the null model for model selection based on [X/Fe] for consistency with the Bayesian evidence. 

\subsubsection{GCE models\label{gce}}

\par
We tested two sets of GCE models derived from observational data. First, we adopted the empirical GCE model from \citet{bedell_chemical_2018}, which is based on the linear correlations between [X/Fe] and stellar age, 
\begin{equation}
G_{\text{[X/Fe]}}(\mathit{age}) = \Delta\mathit{age} \cdot k_X.
\end{equation}
We adopted the slope $k_X$ derived by \citet{bedell_chemical_2018} or \citet{sun_planets_2025} for the Bedell or PASTA sample, respectively. Whether to use the solar baseline or not, \(\Delta \mathit{age}\) is, theoretically, always the age difference relative to the baseline star, though we need not know the exact age of the hypothetical ``average star''. As the age difference matters more than the exact age, the model only considers the slope \(k_X\), and the intercept is implicitly absorbed into the choice of the baseline.

\par
However, in practice, it is difficult to determine the age of the ``average star'' and the prior distribution of \(\Delta \mathit{age}\). So we set the age of the baseline star to be solar age and fit for the stellar age (\(\Delta \mathit{age} = \mathit{age} - \mathit{age}_\odot\)), instead.

\par
When the modeling is based on [X/H], we simply add [Fe/H] to the model prediction:
\begin{equation}
G_{\text{[X/H]}}(\mathit{age}) = \Delta\mathit{age} \cdot k_X + \Delta\text{[Fe/H]}.
\end{equation}
Here \(\Delta\text{[Fe/H]} = \rm [Fe/H] - [Fe/H]_{baseline}\). This means the model will treat $\Delta$[Fe/H] for any star as a known parameter, thus removing Fe from the fitting. 

\par
Second, we also adopted a two-process GCE model, which describes stellar abundance as a combination of a prompt process corresponding to chemical enrichment by massive stars and core-collapse supernovae, and a delayed enrichment process by Type Ia supernovae \citep{griffith_abundance_2019,weinberg_chemical_2019,griffith_residual_2022,weinberg_chemical_2022}. The outcome of these two processes is described as:
\begin{equation}
G_{\text{[X/H]}}(A_{\text{cc}}, A_{\text{Ia}}) = \log_{10} (A_{\text{cc}}q_{\text{cc},X} + A_{\text{Ia}}q_{\text{Ia},X}).
\end{equation}
The coefficients \(q_{i,X}\) for the element \(X\) are determined by the [Mg/H] of the given star, and the amplitudes \(A_i\) for a given star are the same for all elements. We use the correlations derived from the GALAH DR3 sample by \citet{griffith_residual_2022} to determine the coefficients \(q_{i,X}\) from [Mg/H], which means they are not available after changing the baseline, as they were derived based on [X/H]. 

\par
When using [X/Fe], the model becomes $G_{\text{[X/H]}} - G_{\text{[Fe/H]}}$, and thus the amplitudes $A_{\text{Ia}}$ and $A_{\text{cc}}$ become degenerate; therefore, we adopt their ratio \(r = A_{\text{cc}}/ A_{\text{Ia}}\) as the free parameter:
\begin{equation}
    G_{\text{[X/Fe]}} = \log_{10} \displaystyle 
\frac{q_{\text{Ia}, X} + rq_{\text{cc},X}}{q_{\text{Ia, Fe}} +  r q_{\text{cc, Fe}}}.
\end{equation}
Not all elements in our dataset have coefficients available from \citet{griffith_residual_2022} --- only for O, Si, Ca, Ti, Na, Al, K, Sc, Cr, Mn, Fe, Ni, Cu, and Zn; thus, we only fit for these elements for the GALAH GCE model.

\subsubsection{Planet engulfment models}

The engulfment model (E) we use is the same as that in previous works \citep{chambers_slar_2010, oh_kronos_2018, behmard_planet_2023, liu_at_2024}. This model assumes that the materials from the engulfed planet are simply diluted in the star's convection zone. If the materials are mixed immediately after the engulfment, then we have Equation \ref{eqn:raw_engulf}.
With the assumption that a star has the same abundances as the Sun before the engulfment (i.e., solar baseline), \(f_{X,\odot}\) in this equation represents the elemental mass fraction of the Sun \citep{asplund_chemical_2009}. When the baseline is switched to the sample mean, \(f_{X,\odot}\) should represent the elemental mass fraction of the hypothetical ``average star''. 

\par
We note that $M_p$ and $f_{\rm cz}$ are two degenerate parameters, and we thus parametrize engulfment as 
\begin{equation}
    M_{\rm p,eq} = M_{\rm p}M_\odot f_{\rm cz, \odot}/ M_{\text{star}}f_{\rm cz},
\end{equation} 
instead of assigning $f_{\rm cz}$ to each star in our sample (see Section \ref{discussion} for more details). In this work, \(M_\odot f_{\rm cz, \odot}\) is fixed to be \(7000M_\oplus\). Combining Equation \ref{eqn:raw_engulf} and Equation \ref{eqn:meq} gives

\begin{equation}
E_{\rm [X/H]} = \log_{10} \left[1+
\frac{M_{\rm p,eq}f_{X,\rm planet}}
{M_\odot f_{{\rm cz},\odot}f_{X,\odot}}  \right],
\label{eqn:engulfment}
\end{equation}

which is the final form used in the model.

\par
When we use [X/Fe] instead of [X/H], the engulfment model becomes 
\begin{equation}
E_{\text{[X/Fe]}}(M_{\rm p,eq}) = E_{\text{[X/H]}}(M_{\rm p,eq}) - E_{\text{[Fe/H]}}(M_{\rm p,eq}).
\end{equation}

\par
We also introduced an offset to the predicted abundance differences to take into account the potential overall metallicity difference between the star and the Sun, which gives the engulfment model with offset \(E_{\rm offset}\):
\begin{equation}
    E_{\rm offset} = E + \mathit{offset}.
\end{equation}
Although under our fiducial model setup (Sun-[X/Fe]), the effect of atomic diffusion is reduced; in [X/H] cases, adding an offset to the engulfment model is helpful to mitigate the effects of atomic diffusion. This is why E\(_{\rm offset}\) is not used for [X/Fe] but for [X/H] setups.

\par
Under [X/H] setups, the E\(_{\rm offset}\) model is thought to be more general than E, as it may select metal-poor and metal-rich engulfment candidates. However, in reality, a star passing the E - F test may not pass the E\(_{\rm offset}\) - F test, as the model with the offset has an extra parameter and is less competitive in model comparison. Using [X/Fe] is also good for selecting metal-poor and metal-rich engulfment candidates. However, some candidates selected by E\(_{\rm offset}\) under [X/H] can be left out by E under [X/Fe], as their [Fe/H] is not well fitted by E\(_{\rm offset}\).

\subsection{Prior distributions\label{app:prior}}

\par
We made the prior distributions of the free parameters of each model as uninformative as possible, as shown in Table \ref{tab:prior}. The ranges of the uniform distributions are ensured to be slightly larger than the range of the best-fitting parameters in the sample. The distribution of the amplitude ratio \(r\) results from the uniform distributions of \(A_{\text{cc}}/(A_{\text{cc}} + A_{\text{Ia}})\) and \(A_{\text{Ia}}/(A_{\text{cc}} + A_{\text{Ia}})\).

\subsection{Complete matrix of model comparisons\label{app:test}}

The complete matrix of model fitting in this work is listed in Table~\ref{tab:test}.

\begin{deluxetable}{ccccc}[h]
\tablecaption{The Bayesian tests we conduct}\label{tab:test}
\tablehead{
\colhead{Models} & \multicolumn{2}{c}{Baseline: the Sun} & \multicolumn{2}{c}{Baseline: sample mean} \\
{} & \colhead{[X/H]} & \colhead{[X/Fe]} & \colhead{[X/H]} & \colhead{[X/Fe]}
}
\startdata
Flat (F) & \checkmark & \checkmark & \checkmark & \checkmark\\
Engulfment, bulk earth & \checkmark & \checkmark & \checkmark & \checkmark\\
Engulfment with offset & \checkmark &   & \checkmark &  \\
Engulfment, CM chondrite & \checkmark & \checkmark & \checkmark & \checkmark \\
Engulfment with offset, CM chondrite & \checkmark &   & \checkmark &  \\
GCE (Bedell) & \checkmark & \checkmark & \checkmark & \checkmark\\
GCE (GALAH) & \checkmark & \checkmark &  & \\
\enddata
\label{tab:test}
\end{deluxetable}

In our fiducial setup (Sun-[X/Fe]), a star is classified as an engulfment candidate if either the bulk-Earth or CM-chondrite model exceeds its calibrated threshold relative to all three alternatives: $F$, $G_{\rm Bedell}$, and $G_{\rm GALAH}$. With the sample mean as the baseline, however, $G_{\rm GALAH}$ is not applicable. In addition, when using [X/H], an engulfment candidate also needs to pass the criteria for \(\Delta \ln Z (E_{\rm offset} - F)\).

\section{Mock signals and cutoffs\label{mock_signals}}

\begin{figure*}
    \centering
    \includegraphics[width=0.95\linewidth]{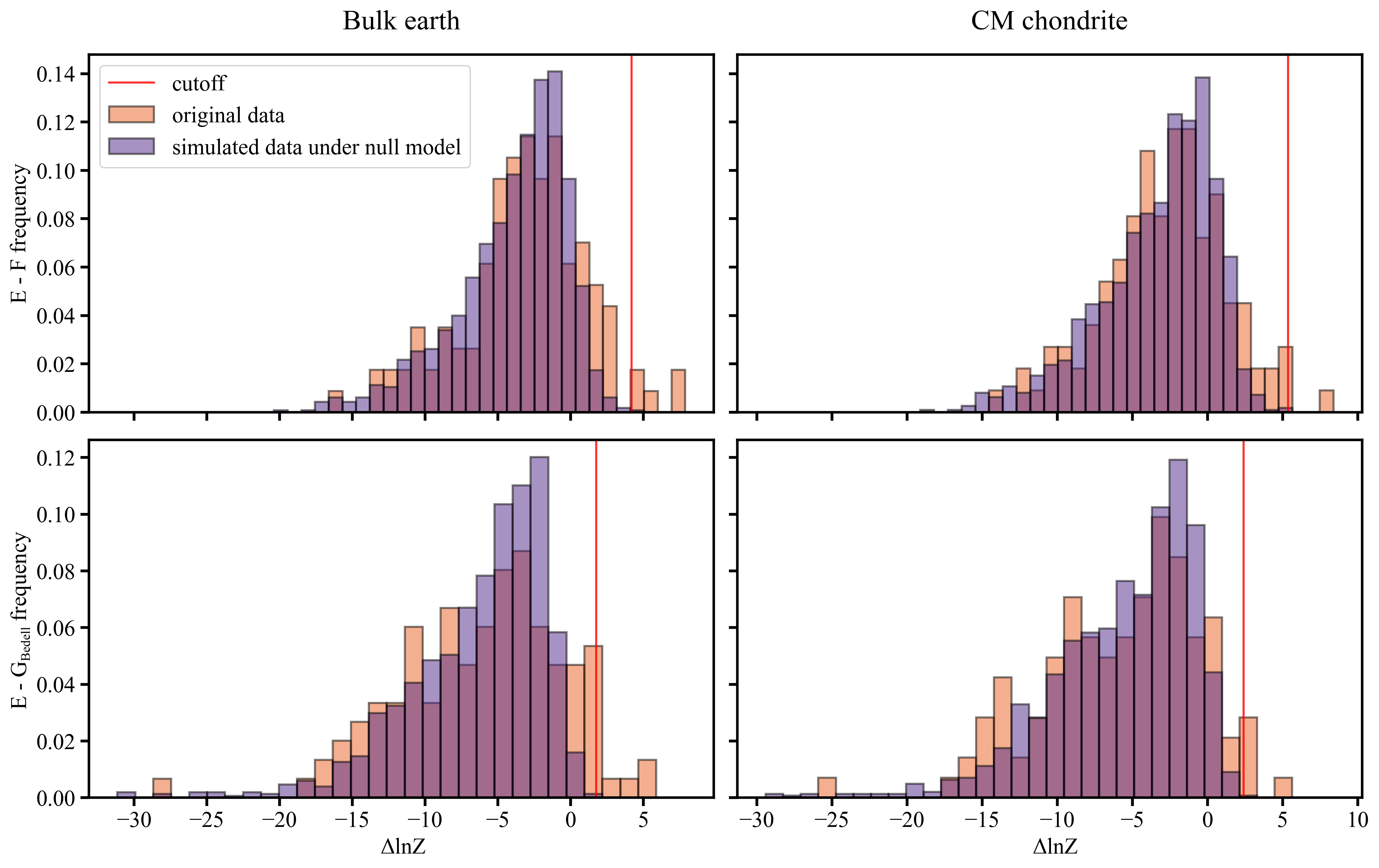}
    \caption{Mock signals and cutoffs under Mean-[X/Fe]. The complete figure set (3 images) of all the mock-test results is available in the online journal.}
    \label{fig:mock_mean_xfe}
\end{figure*}

The mock signals and corresponding cutoffs under other abundance setups (in addition to Sun-[X/Fe]) are shown in Fig.~Set.~\ref{fig:mock_mean_xfe}.

\figsetstart
\figsetnum{\ref{fig:mock_mean_xfe}}
\figsettitle{Full results of the mock tests}
\figsetgrpstart
\figsetgrpnum{\ref{fig:mock_mean_xfe}.1}
\figsetgrptitle{Mean-[X/Fe]}
\figsetplot{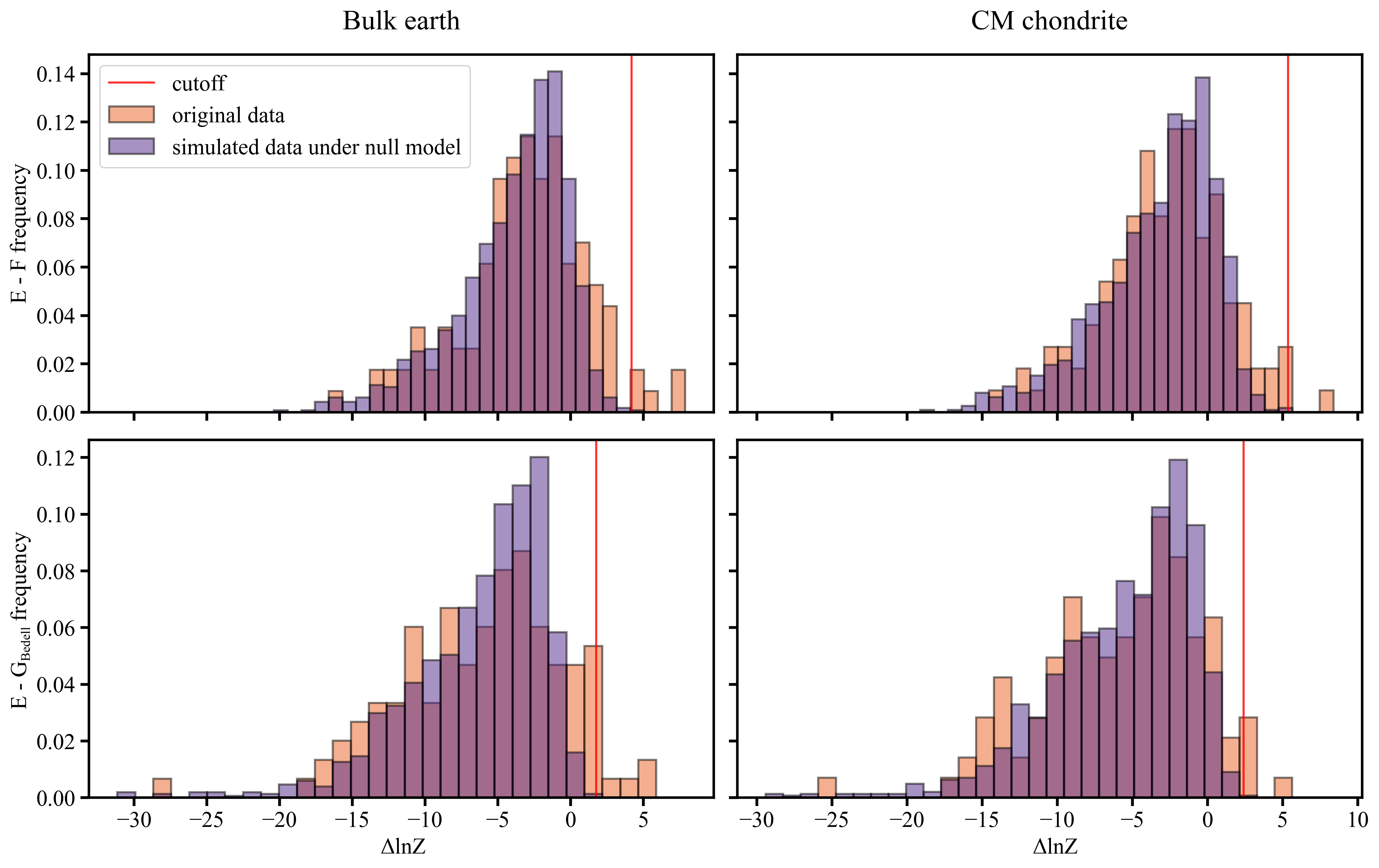}
\figsetgrpnote{Mock signals and cutoffs under Mean-[X/Fe].}
\figsetgrpend
\figsetgrpstart
\figsetgrpnum{\ref{fig:mock_mean_xfe}.2}
\figsetgrptitle{Sun-[X/H]}
\figsetplot{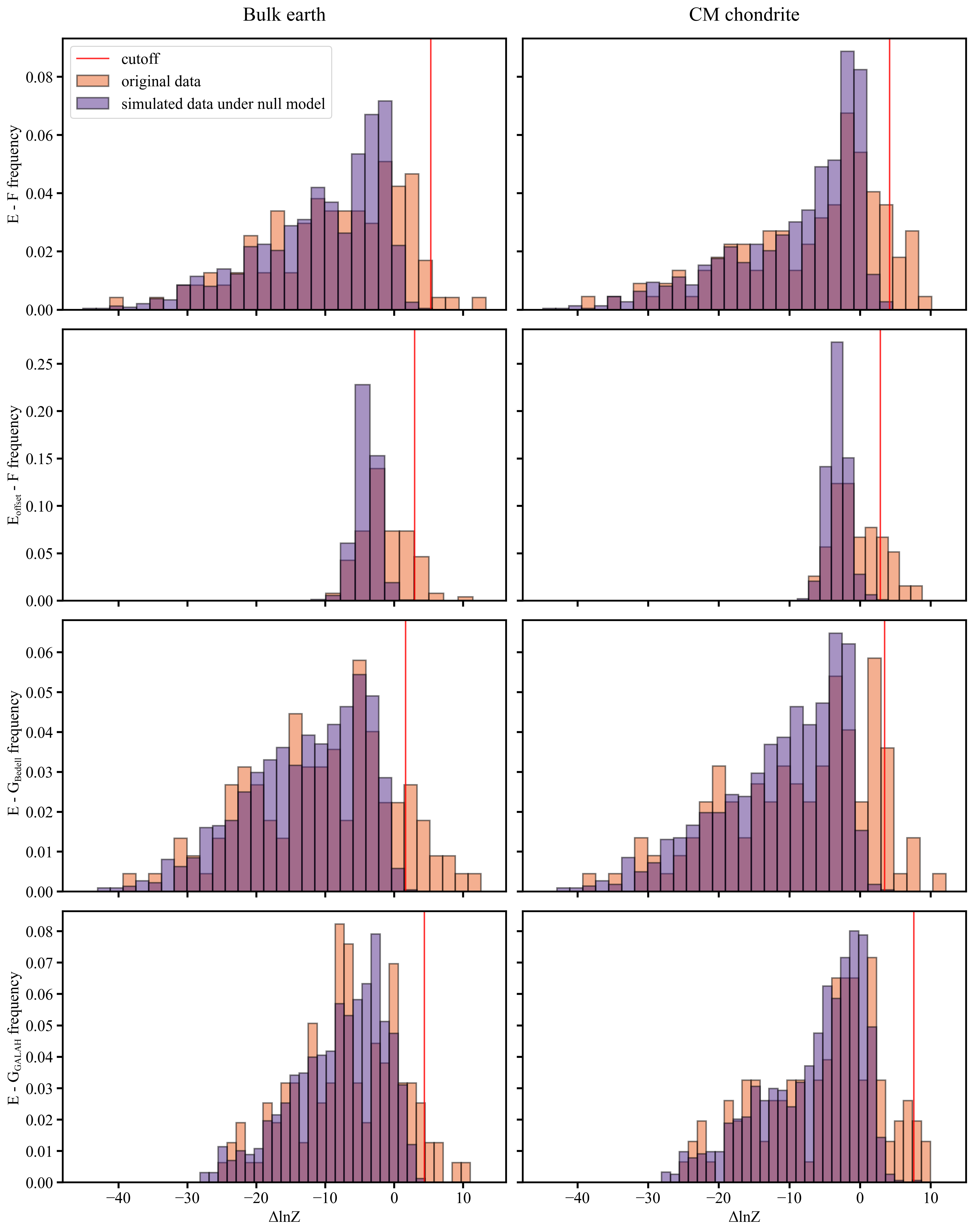}
\figsetgrpnote{Mock signals and cutoffs under Sun-[X/H].}
\figsetgrpend
\figsetgrpstart
\figsetgrpnum{\ref{fig:mock_mean_xfe}.3}
\figsetgrptitle{Mean-[X/H]}
\figsetplot{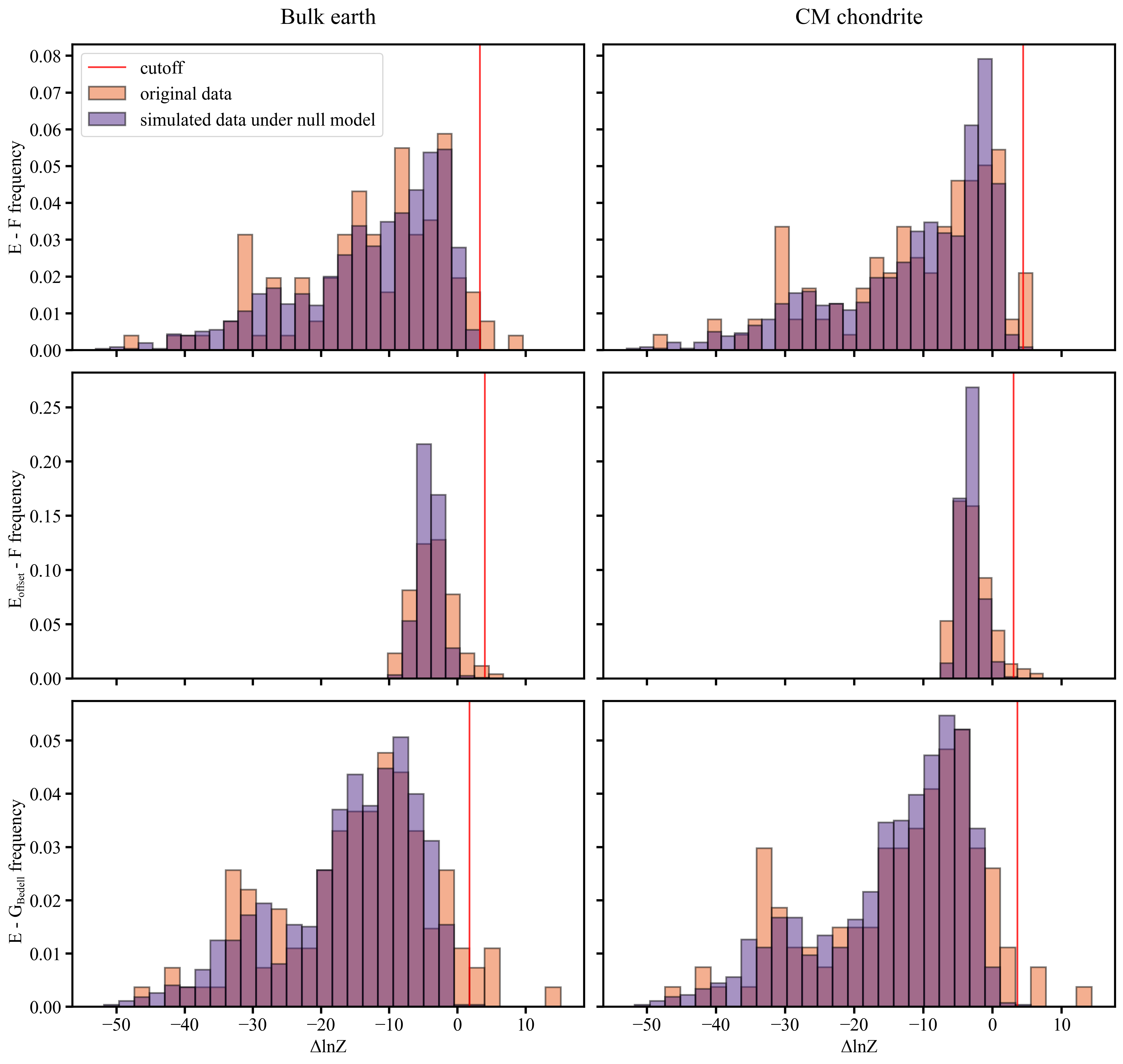}
\figsetgrpnote{Mock signals and cutoffs under Mean-[X/H].}
\figsetgrpend
\figsetend

\section{Results of the mirror test\label{mirror}}

\par
The engulfment signatures have a unidirectional nature. If the baseline is a star that has not engulfed any planet, none of the stars in the sample should show a reversed engulfment pattern (i.e., no stars can have a negative engulfed mass). The mirror test uses this unidirectional nature to check if the baseline is a non-engulfment star. If the distribution of \(\Delta \ln Z (\text{Engulfment} - \text{null})\) of the mirror group shows no significant difference compared to the original data, the baseline is not a non-engulfment star.

\par
\begin{figure*}
    \centering
    \includegraphics[width=\linewidth]{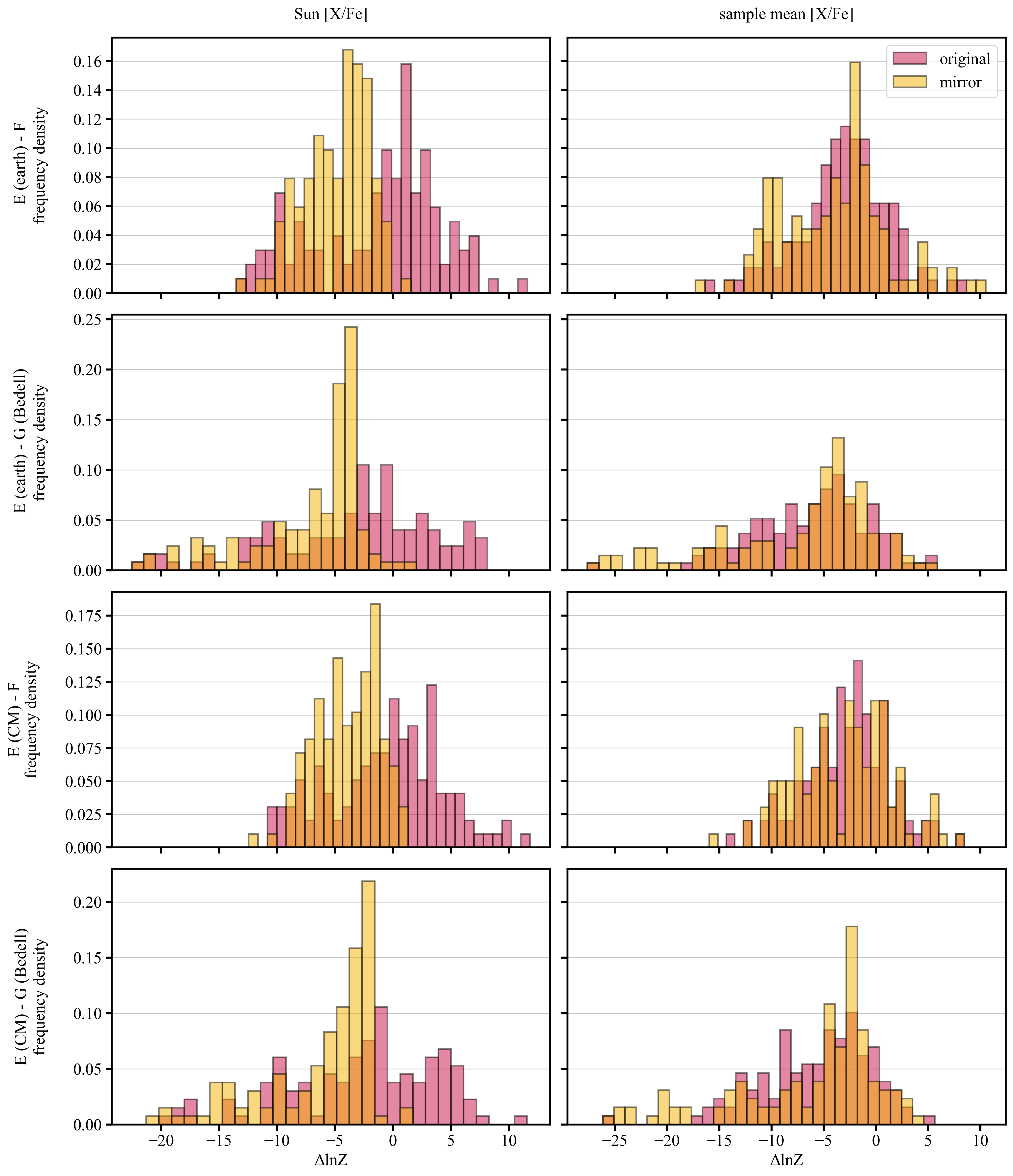}
    \caption{The \(\Delta \ln Z\) distributions of the mirror group (yellow bars) and the original data (magenta bars) for different model comparisons under different baselines. Note that the two distributions in each subplot are almost the same on the sample mean side (the right column), while those on the solar baseline side are not.}
    \label{fig:mirror1}
\end{figure*}
\begin{figure*}
    \centering
    \includegraphics[width=\linewidth]{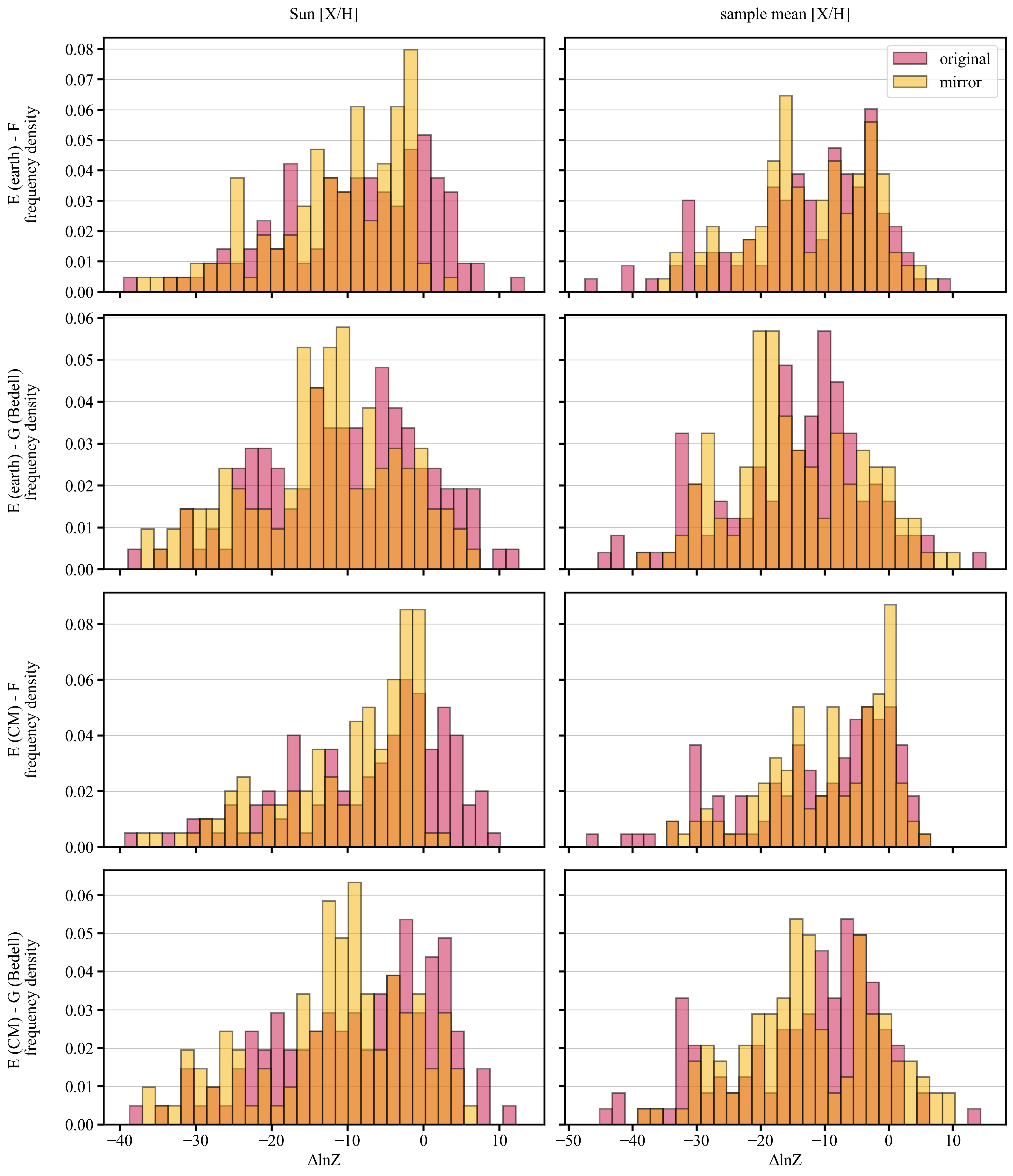}
    \caption{The same description as \ref{fig:mirror1}, except that the abundance type is [X/H] in this comparison.}
    \label{fig:mirror2}
\end{figure*}

The histograms shown in Figure \ref{fig:mirror1} and \ref{fig:mirror2} demonstrate that, compared to the sample mean, the Sun exhibits a non-engulfment nature to some extent. The results of the mirror test indicate that using the Sun as the baseline highlights the engulfment signatures, which is another reason why we chose the solar baseline for our main conclusion. However, the candidates selected under the sample mean baseline are worth attending to. Note that the mirror test is not applicable for the GALAH GCE model, as it is not a simple linear model with good bidirectional nature.

\section{Bayes factors for each test}\label{app:lnz}

\par
The log Bayes factors for each test, \(\Delta \ln Z\), are provided in Table \ref{tab:bulk_earth} and Table \ref{tab:cm_chondrite} (full tables are available in their entirety in a machine-readable form). In addition to the candidates that have met all the criteria, the stars that have partially met the criteria are also worth further investigation, for example, the additional 14 optimistic candidates.

We selected these optimistic candidates using relaxed selection criteria, with \(\ln Z (E - F)\) values exceeding the threshold (under the Sun-[X/Fe] model setup). This is motivated by the fact that the fitting residuals of GCE models \citep[0.01--0.1~dex, ][]{bedell_chemical_2018, griffith_residual_2022} are comparable to or even larger than those of the flat model (0.008--0.03~dex). This more permissive selection includes 13 more stars with bulk Earth composition (11 from Bedell and 2 from PASTA) and 9 stars with CM chondrite composition (8 from Bedell and 1 from PASTA). Among these, 8 stars have passed both Earth and CM selections, including HIP~30502. Among the eight overlapping stars between the Bedell and PASTA samples, only HIP~25670 (TOI-440) was identified as an optimistic candidate (with CM composition) in the Bedell sample, but not within the PASTA sample, most likely due to PASTA's larger error bars in abundances (typically 0.04 dex in PASTA and 0.006 dex in Bedell for this star). Thus, in total, we have 14 additional optimistic candidates who pass this more permissive selection.

\section{Detection Rate Estimates of Planet Engulfment Signatures}\label{app:rate}

Here, we calculate more carefully the detection rate of planet-engulfment signatures in our sample. We estimate, in a very simple manner, the false-positive rate and detection completeness (i.e., false-negative rate) to calculate credible intervals for the detection rate. We also quantify the difference between the detection rate estimates of the Bedell and PASTA samples. Finally, we briefly discuss the potential origins of the engulfed planets and how that compares with some observational evidence at the very end.

Before we begin, we caution that more rigorous estimates of the false-positive and false-negative rates would require injection-recovery simulations across various null-model scenarios and, ideally, cross-validation tests with large samples, which are beyond the scope of this paper. The calculations presented here are meant to provide a quick and simple gauge of how much the reported raw detection rate in the main text would differ from a more careful estimate. 

Consider a sample of size \(N\), with \(n\) detections of engulfment candidates. A simple estimate for the false positive rate \(p\) can be taken as the p-value from our null model simulation test when determining the cutoff value for the log Bayes factor in model comparisons: If any of the null models (flat or GCE) is the underlying truth, then the probability of reporting erroneously a planet engulfment signature detection would be smaller than 1/1210, where 1210 is the sample size for the mocks generated based on null models (Section~\ref{validation_scheme} and Figure~\ref{fig:mock_xfe}). 

The false negative rate \(q\) describes the probability that a star that has engulfed planets is not identified as an engulfment candidate due to limitations in measurement precision and methodology. This is much harder to estimate since the elemental abundance pattern of any star could be a mixture of random scatter, GCE effects, and others, including potential planetary signals. Therefore, typical injection-recovery tests may not fully reflect the true detection sensitivity for individual stars. Here, we adopt some simple assumptions to gauge how the estimated detection rate would change as the adopted false-negative rate varies. 

The first simple assumption is to take \(q = 0\), which assumes that we would have detected the planet-engulfment signature (of masses similar to our best candidates) in any star in our sample if it were truly there, implying 100\% detection completeness. This is convenient, as it provides a conservative estimate of the detection rate; with \(q=0\), the detection rate would be \textit{underestimated}. Another way to obtain a very rough estimate of \(q\) is to assume that, for any given star, if it has a planet-engulfment chemical signature, the significance of detecting such a signature depends solely on the uncertainties in its elemental abundances, or their root-mean-square (RMS). We then compare the RMS of abundance uncertainties of any star with those of our three best candidates: if a star has a lower RMS than a candidate with equivalent engulfed mass $M_{\rm p,eq}$, then we consider the star to have comparable or better sensitivity that would enable the detection of such a chemical signature caused by engulfing $M_{\rm p,eq}$ materials. %This would more likely lead to an underestimation of \(q\) since it ignores other sources of error or mechanisms that could change a star's abundance pattern, in which planet-engulfment signatures could be easily buried --- in another world, the detection sensitivity for any star is probably not as high as suggested by the RMS of abundance errors. This also leads to an \textit{underestimated} detection rate.

To put it all together, from a Bayesian perspective, assuming that the prior distribution of the engulfment rate \(\pi\) is uniform, the posterior distribution for \( \pi \) is

\[
P(\pi|N,n,p,q) \propto (p + (1 - p - q)\pi)^n(1 - p - (1 - q -p)\pi)^{N-n}.
\]

In this work, \(N = 113\), \(n = 3\), and here we adopt \(p =1/1210\) as argued above. Assuming that \(q = 0\), the 95\% credible interval (CI) of the estimated engulfment rate is \([0.88\%,7.42\%]\). When \(q\neq 0\), the estimated engulfment detection rate would increase, as it effectively reduces the sample size. For our 3 engulfment candidates, TOI-3342 has the largest abundance uncertainties, and we take it as our approximate ``detection limit''. The RMS of its abundance uncertainties is 0.030 dex, with 35 sets of abundance measurements in our sample having higher RMS values; all are from the PASTA sample. Assuming that these 35 stars are all false negatives, this sets \(q = 35/113\). In this case, the 95\% CI of the estimated engulfment rate is \([1.28\%,10.75\%]\). In summary, the estimated detection rate is approximately a couple to a few percent, with a lower bound around \(1\%\).

\begin{figure}
    \centering
    \includegraphics[width=9cm]{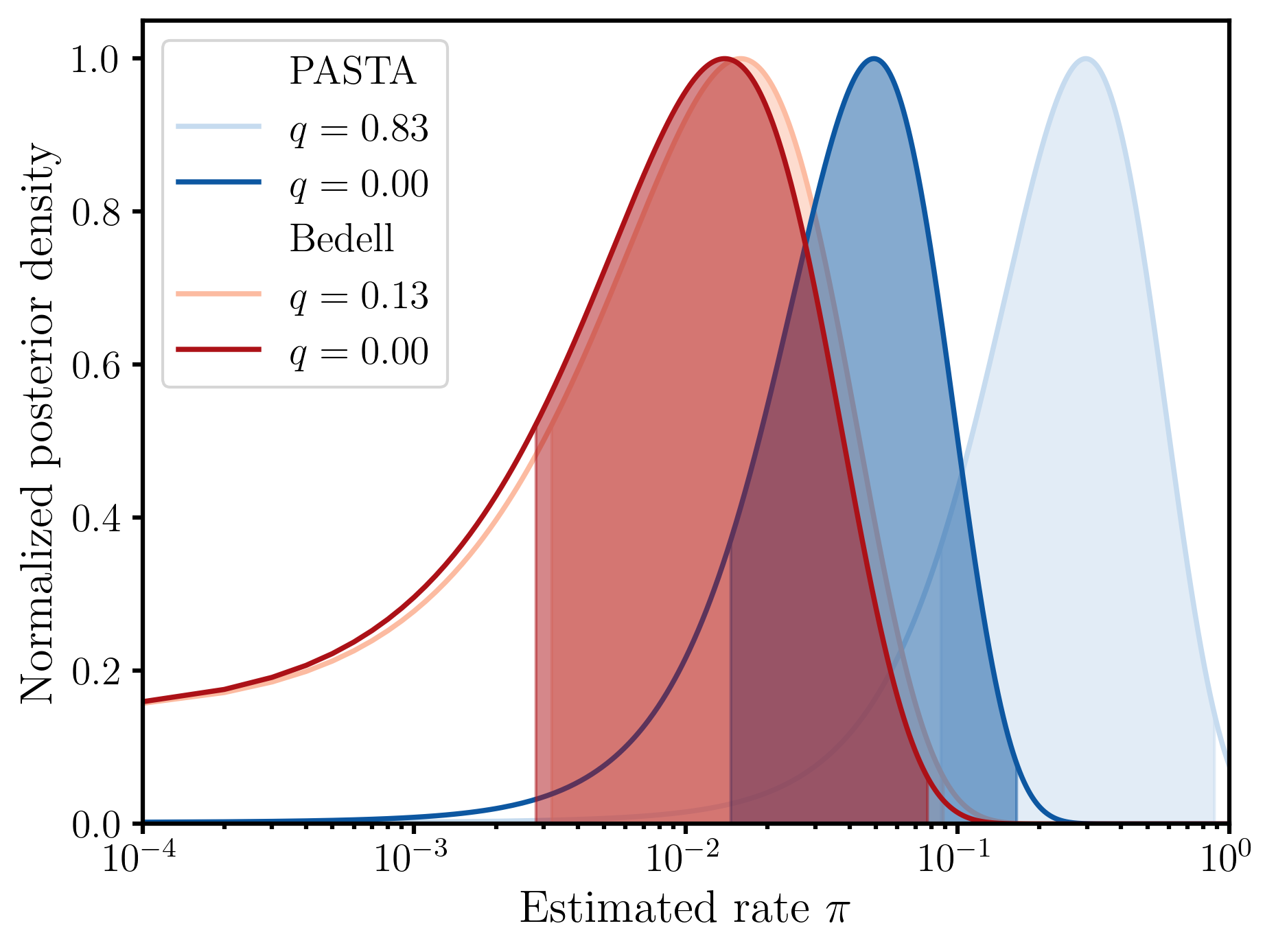}
    \caption{Normalized posterior distribution (rescaling each peak to 1.0) for the estimated detection rate of planet-engulfment signatures in the Bedell (red) and PASTA samples (blue), respectively. Shaded regions are 95\% credible intervals, and dark and light shades indicate results with different assumed false-negative rates or survey completeness, with the darker shade highlighting the extreme case where the false-negative rate is assumed to be 0.00, thus providing a lower-bound estimate.}
    \label{fig:rate}
\end{figure}

We stated earlier that there is no significant difference in engulfment detection rates between the Bedell (mostly non-planet hosts) and the PASTA (planet hosts) samples, given the small numbers of detections in each sample (Section~\ref{diss:rate}). This is also evident from a more careful Bayesian analysis like above. As shown in Figure~\ref{fig:rate}, although the raw rates differ (1/68 for non-planet hosts and 2/40 for planet hosts), assuming \(q = 0\) (i.e., 100\% detection completeness), the 95\% CIs based on the two samples largely overlap. If we consider all stars in the PASTA sample with higher measurement uncertainties (as quantified by the RMS of abundance uncertainties) than TOI-3342 to be false negatives, which gives an extremely large \(q\) of \(35/42 = 0.83\), then the 95\% confidence intervals are separated between the two samples (light shaded curves in Figure~\ref{fig:rate}). Here, in a similar fashion, we also estimated the \(q\) for the Bedell sample by considering that the 10 stars with higher measurement uncertainties than HIP 101905 are false negatives, which yields \(q = 10/79 = 0.13\). We note that the estimated engulfed mass of HIP~101905 is significantly lower than that of TOI-3342, indicating that the Bedell sample has a higher detection sensitivity than the PASTA sample, so this comparison is not entirely fair. In addition, because this estimation of \(q\) is very crude, we need more high-quality data to verify the difference between these two samples.

Mixing the two samples would naturally make the correction ($q$ value estimate) more complicated, but we can argue that, with proper correction, the overall rate could be between the two sets of values presented in Figure~\ref{fig:rate}, assuming that the planet-host and general samples share the same rate (since we see no evidence to the contrary). This means a conservative estimate of the 95\%-CI lower bound for the detection rate is somewhere between 0.3\% and 1.5\% (for the $q=0$ cases, dark-shaded curves in Figure~\ref{fig:rate}), with a basically unconstrained 95\%-CI upper bound of up to 88\% (the light-shaded curves in Figure~\ref{fig:rate}), largely set by the small number of detections and the large measurement uncertainties in the PASTA sample. Therefore, we consider the lower limits of the detection rate to be more meaningful.

The Bedell sample has higher precision than PASTA and represents a generic sample of solar twins and analogs, which is more-or-less brightness-limited and a natural mix of planet and non-planet hosts. Therefore, we could take the detection rate estimate based on this sample to be the best estimate for the general population of solar twins and analogs, which is \([0.3\%,8\%]\) (red curves in Figure~\ref{fig:rate}) for equivalent engulfed masses $\geq 8.2 M_\oplus$ (as set by HIP~101905). This is basically consistent with our nominal estimate based on the raw rates (1--3\%) as presented in the main text. We thus chose to present the raw estimate as the main conclusion, as it is more direct and simpler and carries fewer assumptions.

% Matching detection rate with the occurrence rate of planet engulfment events 
Finally, we emphasize that the rates reported in this work should be interpreted as \textit{detection rates of chemical signatures} of planet engulfment, rather than as the intrinsic occurrence rate of \textit{engulfment events}. A simple duty-cycle estimate illustrates this distinction. If a planet-hosting star loses one planet over $\sim 7$~Gyr, motivated by the decline in the average number of \textit{Kepler}-like planets per system with age from \citet{Yang2023}, and if half of the lost planets are engulfed by the host star, then the expected fraction of planet-hosting stars with visible engulfment signatures is approximately $0.5\tau_{\rm sig}/7~{\rm Gyr}$. For the $\sim 20$~Myr--1.7~Gyr refractory-signature lifetimes estimated by \citet{behmard_planet_2023}, this corresponds to $\sim 0.1$--10\%, broadly encompassing our raw detection fraction of 2/40. However, given the simplicity of this estimate and the uncertainties in the fate of dynamically lost planets, the engulfed masses, the visibility timescale, and our detection completeness, this comparison only illustrates that our estimated detection rate is broadly plausible given the current understanding of planet engulfment events, which motivates larger homogeneous samples to calibrate this connection.

\begin{deluxetable}{cccc}
\tablecaption{The prior distributions of all the models used in this work}\label{tab:prior}
\tablehead{
   \colhead{Parameter} & \colhead{Unit} & \colhead{Description} & \colhead{Prior}
}
\startdata
\multicolumn{4}{l}{Flat} \\
\(\mathit{offset}\) & dex & the overall abundance difference & \(\mathcal{U}(-0.5,0.5)\)\\ \hline
\multicolumn{3}{l}{Engulfment} \\
\(M_{\rm p,eq}\) & \(M_\oplus\) & the equivalent engulfed mass & \(\mathcal{U}(0.5,80)\)\\
\(\mathit{offset}\) & dex & the offset for Eo models & \(\mathcal{U}(-0.5,0.5)\)\\ \hline
\multicolumn{4}{l}{Bedell GCE} \\
\(\mathit{age}\) & Gyr & the stellar age & \(\mathcal{U}(0,12)\)\\ \hline
\multicolumn{4}{l}{GALAH GCE ([X/H])} \\
\(A_{\text{cc}}\) & \nodata & the contribution amplitude of core-collapse supernova  & \(\mathcal{U}(0,3)\)\\
\(A_{\text{Ia}}\) & \nodata & the contribution amplitude of Ia supernova & \(\mathcal{U}(0,3)\)\\ \hline
\multicolumn{4}{l}{GALAH GCE ([X/Fe])} \\
\(r\) & \nodata & the ratio of \(A_{\text{cc}}\) and \(A_{\text{Ia}}\)  & \(P(r) = (1+r)^{-2},r > 0\) \\ \hline
\multicolumn{4}{l}{Intrinsic scatter}\\
\(\sigma_{\text{int}}\) & dex & the intrinsic scatter applied to all the models & \(\mathcal{U}(0,1)\)
\enddata

\end{deluxetable}

\newpage
\begin{deluxetable*}{cc|cccc|ccc|ccc|cc}
\tablecaption{The Log Bayesian evidence for each test (bulk earth)}\label{tab:bulk_earth}

\tablehead{
\multicolumn{2}{c}{} & \multicolumn{7}{c}{Baseline: the Sun} & \multicolumn{5}{c}{Baseline: sample mean} \\
\multicolumn{2}{c}{} & \multicolumn{4}{c}{[X/H]} & \multicolumn{3}{c}{[X/Fe]} & \multicolumn{3}{c}{[X/H]} & \multicolumn{2}{c}{[X/Fe]} \\
\colhead{Name} & \colhead{Source} & \colhead{\(\text{E} - \text{F}\)} & \colhead{\(\text{E}^\prime - \text{Gb}\)} & \colhead{\(\text{E}^{\prime\prime} - \text{Gg}\)} & \colhead{\(\text{Eo} - \text{F}\)} & \colhead{\(\text{E} - \text{F}\)} & \colhead{\(\text{E} - \text{Gb}\)} & \colhead{\(\text{E}^{\prime\prime} - \text{Gg}\)} & \colhead{\(\text{E} - \text{F}\)} & \colhead{\(\text{E}^\prime - \text{Gb}\)} & \colhead{\(\text{Eo} - \text{F}\)} & \colhead{\(\text{E} - \text{F}\)} & \colhead{\(\text{E} - \text{Gb}\)} \\
}
\startdata
\multicolumn{2}{c|}{\textbf{cutoff (bulk earth)}} & \textbf{5.3} & \textbf{1.6} & \textbf{4.4} & \textbf{3.0} & \textbf{4.3} & \textbf{5.3} & {\textbf{3.0}} & {\textbf{3.3}} & {\textbf{1.8}} & {\textbf{4.0}} & {\textbf{4.2}} & {\textbf{1.8}} \\
\hline
    HIP101905   & Bedell\_2018 & \cellcolor[rgb]{ .573,  .816,  .314}13.29 & \cellcolor[rgb]{ .573,  .816,  .314}7.36 & \cellcolor[rgb]{ .573,  .816,  .314}11.01 & \cellcolor[rgb]{ .573,  .816,  .314}11.41 & \cellcolor[rgb]{ .573,  .816,  .314}11.61 & \cellcolor[rgb]{ .573,  .816,  .314}6.62 & \cellcolor[rgb]{ .573,  .816,  .314}9.34 & \cellcolor[rgb]{ .573,  .816,  .314}9.62 & \cellcolor[rgb]{ .573,  .816,  .314}4.02 & \cellcolor[rgb]{ .573,  .816,  .314}6.73 & \cellcolor[rgb]{ .573,  .816,  .314}4.85 & 0.28 \\
    HIP77052    & Bedell\_2018 & \cellcolor[rgb]{ .573,  .816,  .314}6.71 & 0.25  & \cellcolor[rgb]{ .573,  .816,  .314}8.79 & \cellcolor[rgb]{ .573,  .816,  .314}5.66 & \cellcolor[rgb]{ .573,  .816,  .314}8.87 & 2.98  & \cellcolor[rgb]{ .573,  .816,  .314}7.68 & \cellcolor[rgb]{ .573,  .816,  .314}5.01 & -4.01 & 3.13  & \cellcolor[rgb]{ .573,  .816,  .314}5.62 & -2.90 \\
    TOI-3342 & PASTA\_I & 3.93  & \cellcolor[rgb]{ .573,  .816,  .314}4.42 & 3.93  & \cellcolor[rgb]{ .573,  .816,  .314}4.27 & \cellcolor[rgb]{ .573,  .816,  .314}5.95 & \cellcolor[rgb]{ .573,  .816,  .314}6.85 & \cellcolor[rgb]{ .573,  .816,  .314}4.24 & 2.43  & -0.47 & 1.81  & \cellcolor[rgb]{ .573,  .816,  .314}4.39 & 1.71 \\
    HIP30502    & Bedell\_2018 & -17.67 & -19.26 & -12.71 & \cellcolor[rgb]{ .573,  .816,  .314}4.78 & \cellcolor[rgb]{ .573,  .816,  .314}7.19 & 4.64  & 0.56  & -26.12 & -26.98 & \cellcolor[rgb]{ .573,  .816,  .314}4.42 & \cellcolor[rgb]{ .573,  .816,  .314}7.87 & \cellcolor[rgb]{ .573,  .816,  .314}5.12 \\
    HIP43297    & Bedell\_2018 & \cellcolor[rgb]{ .573,  .816,  .314}7.60 & -0.47 & \cellcolor[rgb]{ .573,  .816,  .314}7.04 & \cellcolor[rgb]{ .573,  .816,  .314}4.15 & \cellcolor[rgb]{ .573,  .816,  .314}5.52 & -1.93 & \cellcolor[rgb]{ .573,  .816,  .314}5.67 & 3.27  & -4.90 & 0.27  & -0.08 & -7.83 \\
    HIP74389    & Bedell\_2018 & 3.67  & -4.65 & \cellcolor[rgb]{ .573,  .816,  .314}6.96 & \cellcolor[rgb]{ .573,  .816,  .314}5.57 & \cellcolor[rgb]{ .573,  .816,  .314}7.19 & -0.50 & \cellcolor[rgb]{ .573,  .816,  .314}10.44 & -1.88 & -10.30 & -0.60 & -1.42 & -9.55 \\
    HIP8507     & Bedell\_2018 & -23.11 & -25.16 & -17.00 & \cellcolor[rgb]{ .573,  .816,  .314}3.20 & \cellcolor[rgb]{ .573,  .816,  .314}6.52 & 2.89  & -0.26 & -31.15 & -36.44 & \cellcolor[rgb]{ .573,  .816,  .314}4.39 & \cellcolor[rgb]{ .573,  .816,  .314}7.63 & 0.26 \\
    HD 75302 & PASTA\_II & 1.53  & \cellcolor[rgb]{ .573,  .816,  .314}6.00 & 0.02  & 2.58  & 2.86  & \cellcolor[rgb]{ .573,  .816,  .314}7.77 & -2.14 & -1.55 & \cellcolor[rgb]{ .573,  .816,  .314}2.61 & 1.29  & 1.13  & \cellcolor[rgb]{ .573,  .816,  .314}5.88 \\
    TOI-248 & PASTA\_II & 4.77  & \cellcolor[rgb]{ .573,  .816,  .314}9.21 & 3.93  & 1.93  & 1.52  & \cellcolor[rgb]{ .573,  .816,  .314}6.59 & 1.06  & \cellcolor[rgb]{ .573,  .816,  .314}3.55 & \cellcolor[rgb]{ .573,  .816,  .314}3.39 & 0.19  & 0.13  & 0.35 \\
    TOI-5005 & PASTA\_I & 2.82  & \cellcolor[rgb]{ .573,  .816,  .314}12.57 & 2.41  & 1.58  & -3.91 & \cellcolor[rgb]{ .573,  .816,  .314}6.04 & 0.53  & 1.87  & \cellcolor[rgb]{ .573,  .816,  .314}15.17 & -0.06 & -9.17 & \cellcolor[rgb]{ .573,  .816,  .314}3.99 \\
    HIP38072    & Bedell\_2018 & 1.87  & -3.35 & 2.89  & \cellcolor[rgb]{ .573,  .816,  .314}3.32 & \cellcolor[rgb]{ .573,  .816,  .314}7.30 & 1.85  & \cellcolor[rgb]{ .573,  .816,  .314}5.72 & -2.36 & -10.50 & -0.61 & 2.67  & -5.55 \\
    HIP25670    & Bedell\_2018 & 0.55  & -4.24 & 1.75  & \cellcolor[rgb]{ .573,  .816,  .314}3.09 & \cellcolor[rgb]{ .573,  .816,  .314}7.02 & 2.38  & \cellcolor[rgb]{ .573,  .816,  .314}10.35 & -3.16 & -7.37 & -2.63 & 0.34  & -3.53 \\
    TOI-2426 & PASTA\_I & -6.38 & -5.93 & -2.52 & \cellcolor[rgb]{ .573,  .816,  .314}3.46 & \cellcolor[rgb]{ .573,  .816,  .314}6.38 & \cellcolor[rgb]{ .573,  .816,  .314}6.12 & 2.78  & -13.61 & -13.93 & 0.89  & 1.15  & -0.28 \\
    HIP105184   & Bedell\_2018 & -4.86 & -11.31 & -1.16 & \cellcolor[rgb]{ .573,  .816,  .314}3.10 & \cellcolor[rgb]{ .573,  .816,  .314}5.73 & -0.72 & \cellcolor[rgb]{ .573,  .816,  .314}3.81 & -8.89 & -21.74 & 1.97  & 1.66  & -11.24 \\
    HIP79672    & Bedell\_2018 & 3.12  & -2.87 & \cellcolor[rgb]{ .573,  .816,  .314}4.69 & 1.03  & \cellcolor[rgb]{ .573,  .816,  .314}4.99 & -0.41 & \cellcolor[rgb]{ .573,  .816,  .314}7.37 & -1.97 & -7.60 & -4.06 & -2.39 & -7.91 \\
    TOI-2479 & PASTA\_II & -6.22 & -2.36 & -2.74 & 2.81  & \cellcolor[rgb]{ .573,  .816,  .314}4.48 & \cellcolor[rgb]{ .573,  .816,  .314}8.15 & 1.30  & -18.08 & -16.78 & -0.32 & 2.75  & \cellcolor[rgb]{ .573,  .816,  .314}2.59 \\
    HIP7585     & Bedell\_2018 & -6.58 & -14.55 & -0.01 & 2.02  & \cellcolor[rgb]{ .573,  .816,  .314}5.20 & -2.63 & \cellcolor[rgb]{ .573,  .816,  .314}10.01 & -8.33 & -14.38 & -3.35 & -2.11 & -8.06 \\
    HIP104045   & Bedell\_2018 & -0.75 & -7.97 & 4.04  & \cellcolor[rgb]{ .573,  .816,  .314}3.07 & 4.26  & -2.67 & \cellcolor[rgb]{ .573,  .816,  .314}7.72 & -3.89 & -7.47 & -4.36 & -3.75 & -7.10 \\
    TOI-2523 & PASTA\_II & 3.35  & \cellcolor[rgb]{ .573,  .816,  .314}2.09 & \cellcolor[rgb]{ .573,  .816,  .314}4.90 & 0.77  & 3.76  & 2.54  & 1.27  & -5.58 & -1.96 & -2.83 & -4.95 & -2.37 \\
    HD 88072 & PASTA\_II & -1.91 & \cellcolor[rgb]{ .573,  .816,  .314}1.86 & -0.64 & 2.94  & 3.52  & \cellcolor[rgb]{ .573,  .816,  .314}7.81 & -2.02 & -17.33 & -16.37 & -2.01 & 1.34  & 0.95 \\
    \(\vdots\) & \(\vdots\) & \(\vdots\) & \(\vdots\) & \(\vdots\) & \(\vdots\) & \(\vdots\) & \(\vdots\) & \(\vdots\) & \(\vdots\) & \(\vdots\) & \(\vdots\) & \(\vdots\) & \(\vdots\) \\
\enddata

\tablecomments{
Same as Table \ref{tab:candidate}, but for the whole sample in bulk earth fitting. This table is available in its entirety in a machine-readable form.
}

\end{deluxetable*}

\newpage
\begin{deluxetable*}{cc|cccc|ccc|ccc|cc}
\tablecaption{The Log Bayesian evidence for each test (CM chondrite)}\label{tab:cm_chondrite}

\tablehead{
\multicolumn{2}{c}{} & \multicolumn{7}{c}{Baseline: the Sun} & \multicolumn{5}{c}{Baseline: sample mean} \\
\multicolumn{2}{c}{} & \multicolumn{4}{c}{[X/H]} & \multicolumn{3}{c}{[X/Fe]} & \multicolumn{3}{c}{[X/H]} & \multicolumn{2}{c}{[X/Fe]} \\
\colhead{Name} & \colhead{Source} & \colhead{\(\text{E} - \text{F}\)} & \colhead{\(\text{E}^\prime - \text{Gb}\)} & \colhead{\(\text{E}^{\prime\prime} - \text{Gg}\)} & \colhead{\(\text{Eo} - \text{F}\)} & \colhead{\(\text{E} - \text{F}\)} & \colhead{\(\text{E} - \text{Gb}\)} & \colhead{\(\text{E}^{\prime\prime} - \text{Gg}\)} & \colhead{\(\text{E} - \text{F}\)} & \colhead{\(\text{E}^\prime - \text{Gb}\)} & \colhead{\(\text{Eo} - \text{F}\)} & \colhead{\(\text{E} - \text{F}\)} & \colhead{\(\text{E} - \text{Gb}\)} \\
}
\startdata
\multicolumn{2}{c|}{\textbf{cutoff (CM chondrite)}} & {\textbf{4.2}} & {\textbf{3.5}} & {\textbf{7.6}} & {\textbf{2.9}} & {\textbf{6.0}} & {\textbf{4.2}} & {\textbf{3.9}} & {\textbf{4.5}} & {\textbf{3.6}} & {\textbf{3.0}} & {\textbf{5.4}} & {\textbf{2.4}} \\ \hline
TOI-2426 & PASTA\_I & -6.46 & -6.10 & -2.82 & \cellcolor[rgb]{ .573,  .816,  .314}8.78 & \cellcolor[rgb]{ .573,  .816,  .314}11.85 & \cellcolor[rgb]{ .573,  .816,  .314}11.58 & \cellcolor[rgb]{ .573,  .816,  .314}4.98 & -13.70 & -13.97 & \cellcolor[rgb]{ .573,  .816,  .314}3.20 & 3.92  & \cellcolor[rgb]{ .573,  .816,  .314}2.49 \\
    HIP77052    & Bedell\_2018 & \cellcolor[rgb]{ .573,  .816,  .314}4.75 & -1.56 & 6.38  & \cellcolor[rgb]{ .573,  .816,  .314}7.47 & \cellcolor[rgb]{ .573,  .816,  .314}9.95 & 4.07  & \cellcolor[rgb]{ .573,  .816,  .314}5.75 & 4.02  & -4.66 & \cellcolor[rgb]{ .573,  .816,  .314}4.74 & \cellcolor[rgb]{ .573,  .816,  .314}5.45 & -3.06 \\
    HIP30502    & Bedell\_2018 & -17.62 & -19.15 & -12.90 & \cellcolor[rgb]{ .573,  .816,  .314}6.18 & \cellcolor[rgb]{ .573,  .816,  .314}9.59 & \cellcolor[rgb]{ .573,  .816,  .314}7.04 & 2.11  & -25.93 & -26.92 & \cellcolor[rgb]{ .573,  .816,  .314}7.30 & \cellcolor[rgb]{ .573,  .816,  .314}8.39 & \cellcolor[rgb]{ .573,  .816,  .314}5.64 \\
    HIP79672    & Bedell\_2018 & \cellcolor[rgb]{ .573,  .816,  .314}6.83 & 0.91  & \cellcolor[rgb]{ .573,  .816,  .314}7.81 & \cellcolor[rgb]{ .573,  .816,  .314}3.30 & \cellcolor[rgb]{ .573,  .816,  .314}6.20 & 0.79  & \cellcolor[rgb]{ .573,  .816,  .314}6.85 & 0.69  & -4.68 & -2.68 & -2.05 & -7.58 \\
    HIP7585     & Bedell\_2018 & \cellcolor[rgb]{ .573,  .816,  .314}4.33 & -4.35 & \cellcolor[rgb]{ .573,  .816,  .314}7.94 & \cellcolor[rgb]{ .573,  .816,  .314}5.00 & \cellcolor[rgb]{ .573,  .816,  .314}6.07 & -1.77 & \cellcolor[rgb]{ .573,  .816,  .314}8.66 & -1.80 & -7.97 & -1.69 & -1.94 & -7.90 \\
    HIP101905   & Bedell\_2018 & \cellcolor[rgb]{ .573,  .816,  .314}7.12 & 1.54  & 6.80  & \cellcolor[rgb]{ .573,  .816,  .314}7.71 & 5.28  & 0.30  & \cellcolor[rgb]{ .573,  .816,  .314}5.81 & \cellcolor[rgb]{ .573,  .816,  .314}5.78 & 0.66  & \cellcolor[rgb]{ .573,  .816,  .314}4.70 & 0.34  & -4.23 \\
    HIP38072    & Bedell\_2018 & \cellcolor[rgb]{ .573,  .816,  .314}7.37 & 1.95  & 7.30  & \cellcolor[rgb]{ .573,  .816,  .314}4.52 & \cellcolor[rgb]{ .573,  .816,  .314}8.87 & 3.42  & \cellcolor[rgb]{ .573,  .816,  .314}6.45 & 2.63  & -5.79 & 0.70  & 3.19  & -5.02 \\
    HIP25670    & Bedell\_2018 & \cellcolor[rgb]{ .573,  .816,  .314}7.44 & 2.07  & 7.16  & \cellcolor[rgb]{ .573,  .816,  .314}4.35 & \cellcolor[rgb]{ .573,  .816,  .314}8.53 & 3.89  & \cellcolor[rgb]{ .573,  .816,  .314}9.68 & 1.50  & -2.73 & -1.21 & 0.99  & -2.88 \\
    HIP40133    & Bedell\_2018 & 3.85  & 1.79  & 1.78  & \cellcolor[rgb]{ .573,  .816,  .314}7.15 & \cellcolor[rgb]{ .573,  .816,  .314}7.13 & \cellcolor[rgb]{ .573,  .816,  .314}4.79 & 1.64  & -1.29 & -3.37 & 1.54  & 5.24  & \cellcolor[rgb]{ .573,  .816,  .314}2.82 \\
    HIP74389    & Bedell\_2018 & \cellcolor[rgb]{ .573,  .816,  .314}10.14 & 1.79  & \cellcolor[rgb]{ .573,  .816,  .314}9.96 & \cellcolor[rgb]{ .573,  .816,  .314}6.57 & 4.39  & -3.30 & \cellcolor[rgb]{ .573,  .816,  .314}6.98 & 2.86  & -5.34 & 0.40  & -2.47 & -10.60 \\
    TOI-5005 & PASTA\_I & 2.25  & \cellcolor[rgb]{ .573,  .816,  .314}12.15 & 1.43  & \cellcolor[rgb]{ .573,  .816,  .314}4.03 & -5.04 & \cellcolor[rgb]{ .573,  .816,  .314}4.91 & -1.01 & 1.17  & \cellcolor[rgb]{ .573,  .816,  .314}14.34 & 0.54  & -10.82 & 2.33 \\
    TOI-1117 & PASTA\_I & \cellcolor[rgb]{ .573,  .816,  .314}5.17 & 2.35  & 1.12  & \cellcolor[rgb]{ .573,  .816,  .314}3.52 & \cellcolor[rgb]{ .573,  .816,  .314}6.15 & 3.64  & 1.16  & 1.64  & -0.48 & -0.40 & 0.53  & -1.57 \\
    TOI-215 & PASTA\_I & \cellcolor[rgb]{ .573,  .816,  .314}4.26 & 3.01  & -2.30 & \cellcolor[rgb]{ .573,  .816,  .314}3.28 & 6.00  & \cellcolor[rgb]{ .573,  .816,  .314}4.73 & -0.40 & 1.36  & -3.36 & -0.21 & 3.13  & 0.39 \\
    HIP85042    & Bedell\_2018 & \cellcolor[rgb]{ .573,  .816,  .314}5.54 & 3.28  & 1.86  & \cellcolor[rgb]{ .573,  .816,  .314}3.01 & 3.90  & 1.69  & -2.35 & \cellcolor[rgb]{ .573,  .816,  .314}4.62 & 1.97  & 0.56  & 4.81  & 2.34 \\
    TOI-2518 & PASTA\_II & -1.36 & \cellcolor[rgb]{ .573,  .816,  .314}4.54 & -0.29 & \cellcolor[rgb]{ .573,  .816,  .314}3.17 & -0.31 & \cellcolor[rgb]{ .573,  .816,  .314}5.82 & -1.24 & -14.57 & -10.32 & -2.29 & -5.31 & -1.67 \\
    HIP8507     & Bedell\_2018 & -22.78 & -24.91 & -16.86 & 2.83  & \cellcolor[rgb]{ .573,  .816,  .314}6.37 & 2.74  & -0.65 & -30.89 & -36.31 & \cellcolor[rgb]{ .573,  .816,  .314}3.18 & 4.46  & -2.91 \\
    TOI-3342 & PASTA\_I & 2.35  & 2.83  & 2.36  & \cellcolor[rgb]{ .573,  .816,  .314}4.35 & 4.58  & \cellcolor[rgb]{ .573,  .816,  .314}5.48 & 3.21  & 1.06  & -1.58 & 0.92  & 2.66  & -0.02 \\
    HD 88072 & PASTA\_II & -2.30 & 1.45  & -1.04 & \cellcolor[rgb]{ .573,  .816,  .314}5.00 & 3.73  & \cellcolor[rgb]{ .573,  .816,  .314}8.02 & -2.09 & -17.06 & -16.46 & -1.52 & 1.07  & 0.68 \\
    HIP43297    & Bedell\_2018 & \cellcolor[rgb]{ .573,  .816,  .314}6.59 & -1.19 & 5.76  & \cellcolor[rgb]{ .573,  .816,  .314}4.49 & 3.34  & -4.10 & 3.75  & 4.04  & -4.03 & 1.17  & -1.33 & -9.08 \\
    TOI-2479 & PASTA\_II & -6.31 & -2.58 & -2.87 & \cellcolor[rgb]{ .573,  .816,  .314}3.44 & 3.33  & \cellcolor[rgb]{ .573,  .816,  .314}7.00 & 0.46  & -18.08 & -16.93 & -1.80 & 0.47  & 0.32 \\
    \(\vdots\) & \(\vdots\) & \(\vdots\) & \(\vdots\) & \(\vdots\) & \(\vdots\) & \(\vdots\) & \(\vdots\) & \(\vdots\) & \(\vdots\) & \(\vdots\) & \(\vdots\) & \(\vdots\) & \(\vdots\) \\
\enddata

\tablecomments{
Same as Table \ref{tab:candidate}, but for the whole sample in CM chondrite fitting. This table is available in its entirety in a machine-readable form.
}

\end{deluxetable*}

%% For this sample we use BibTeX plus aasjournalv7.bst to generate the

\bibliography{mybib}{}

@article{cheng_mesa_2026,
	title = {A {MESA} {Grid} of {Convection}-zone {Masses} for {Sun}-like {Main}-sequence {Stars}},
	volume = {10},
	issn = {2515-5172},
	url = {https://iopscience.iop.org/article/10.3847/2515-5172/ae8392},
	doi = {10.3847/2515-5172/ae8392},
	language = {en},
	number = {7},
	urldate = {2026-07-03},
	journal = {Research Notes of the AAS},
	publisher = {IOP Publishing},
	author = {Cheng, Zimo and Sun, Meng and Wang, Sharon X.},
	month = jul,
	year = {2026},
	pages = {180},
}

@ARTICLE{Gonzalez2010,
       author = {{Gonz{\'a}lez Hern{\'a}ndez}, J.~I. and {Israelian}, G. and {Santos}, N.~C. and {Sousa}, S. and {Delgado-Mena}, E. and {Neves}, V. and {Udry}, S.},
        title = "{Searching for the Signatures of Terrestrial Planets in Solar Analogs}",
      journal = {\apj},
         year = 2010,
        month = sep,
       volume = {720},
       number = {2},
        pages = {1592-1602},
          doi = {10.1088/0004-637X/720/2/1592},
archivePrefix = {arXiv},
       eprint = {1007.0580},
 primaryClass = {astro-ph.SR},
       adsurl = {https://ui.adsabs.harvard.edu/abs/2010ApJ...720.1592G}
}

@ARTICLE{Gonzalez2013,
       author = {{Gonz{\'a}lez Hern{\'a}ndez}, J.~I. and {Delgado-Mena}, E. and {Sousa}, S.~G. and {Israelian}, G. and {Santos}, N.~C. and {Adibekyan}, V. Zh. and {Udry}, S.},
        title = "{Searching for the signatures of terrestrial planets in F-, G-type main-sequence stars}",
      journal = {\aap},
         year = 2013,
        month = apr,
       volume = {552},
          eid = {A6},
        pages = {A6},
          doi = {10.1051/0004-6361/201220165},
archivePrefix = {arXiv},
       eprint = {1301.2109},
 primaryClass = {astro-ph.EP},
       adsurl = {https://ui.adsabs.harvard.edu/abs/2013A&A...552A...6G}
}

@ARTICLE{Levrard2009,
       author = {{Levrard}, B. and {Winisdoerffer}, C. and {Chabrier}, G.},
        title = "{Falling Transiting Extrasolar Giant Planets}",
      journal = {\apjl},
         year = 2009,
        month = feb,
       volume = {692},
       number = {1},
        pages = {L9-L13},
          doi = {10.1088/0004-637X/692/1/L9},
archivePrefix = {arXiv},
       eprint = {0901.2048},
 primaryClass = {astro-ph.EP},
       adsurl = {https://ui.adsabs.harvard.edu/abs/2009ApJ...692L...9L}
}

@ARTICLE{Li2014,
       author = {{Li}, Gongjie and {Naoz}, Smadar and {Valsecchi}, Francesca and {Johnson}, John Asher and {Rasio}, Frederic A.},
        title = "{The Dynamics of the Multi-planet System Orbiting Kepler-56}",
      journal = {\apj},
         year = 2014,
        month = oct,
       volume = {794},
       number = {2},
          eid = {131},
        pages = {131},
          doi = {10.1088/0004-637X/794/2/131},
archivePrefix = {arXiv},
       eprint = {1407.2249},
 primaryClass = {astro-ph.EP},
       adsurl = {https://ui.adsabs.harvard.edu/abs/2014ApJ...794..131L}
}

@ARTICLE{Yee2020,
       author = {{Yee}, Samuel W. and {Winn}, Joshua N. and {Knutson}, Heather A. and {Patra}, Kishore C. and {Vissapragada}, Shreyas and {Zhang}, Michael M. and {Holman}, Matthew J. and {Shporer}, Avi and {Wright}, Jason T.},
        title = "{The Orbit of WASP-12b Is Decaying}",
      journal = {\apjl},
         year = 2020,
        month = jan,
       volume = {888},
       number = {1},
          eid = {L5},
        pages = {L5},
          doi = {10.3847/2041-8213/ab5c16},
archivePrefix = {arXiv},
       eprint = {1911.09131},
 primaryClass = {astro-ph.EP},
       adsurl = {https://ui.adsabs.harvard.edu/abs/2020ApJ...888L...5Y}
}

@article{skilling_nested_2004,
	title = {Nested {Sampling}},
	volume = {735},
	issn = {0094-243X},
	url = {https://ui.adsabs.harvard.edu/abs/2004AIPC..735..395S/abstract},
	doi = {10.1063/1.1835238},
	language = {en},
	urldate = {2026-06-27},
	journal = {Bayesian Inference and Maximum Entropy Methods in Science and Engineering: 24th International Workshop on Bayesian Inference and Maximum Entropy Methods in Science and Engineering},
	author = {Skilling, John},
	month = nov,
	year = {2004},
	pages = {395--405},
}

@article{skilling_nested_2006,
	title = {Nested sampling for general {Bayesian} computation},
	volume = {1},
	issn = {1936-0975, 1931-6690},
	url = {https://projecteuclid.org/journals/bayesian-analysis/volume-1/issue-4/Nested-sampling-for-general-Bayesian-computation/10.1214/06-BA127.full},
	doi = {10.1214/06-BA127},
	language = {en},
	number = {4},
	urldate = {2026-06-27},
	journal = {Bayesian Analysis},
	publisher = {International Society for Bayesian Analysis},
	author = {Skilling, John},
	month = dec,
	year = {2006},
	pages = {833--859},
}

@article{feroz_multinest_2009,
	title = {{MULTINEST}: an efficient and robust {Bayesian} inference tool for cosmology and particle physics},
	volume = {398},
	issn = {0035-8711},
	shorttitle = {{MULTINEST}},
	url = {https://ui.adsabs.harvard.edu/abs/2009MNRAS.398.1601F/abstract},
	doi = {10.1111/j.1365-2966.2009.14548.x},
	language = {en},
	number = {4},
	urldate = {2026-06-27},
	journal = {Monthly Notices of the Royal Astronomical Society},
	author = {Feroz, F. and Hobson, M. P. and Bridges, M.},
	month = oct,
	year = {2009},
	pages = {1601--1614},
}

@software{sergey_koposov_2025_17268284,
  author       = {Sergey Koposov and
                  Josh Speagle and
                  Kyle Barbary and
                  Gregory Ashton and
                  Ed Bennett and
                  Johannes Buchner and
                  Carl Scheffler and
                  Colm Talbot and
                  Ben Cook and
                  James Guillochon and
                  Patricio Cubillos and
                  Andrés Asensio Ramos and
                  Matthieu Dartiailh and
                  Ilya and
                  Erik Tollerud and
                  Dustin Lang and
                  Ben Johnson and
                  jtmendel and
                  Edward Higson and
                  Thomas Vandal and
                  Tansu Daylan and
                  Ruth Angus and
                  patelR and
                  Phillip Cargile and
                  Patrick Sheehan and
                  Matt Pitkin and
                  Matthew Kirk and
                  Lu Xu and
                  Joel Leja and
                  joezuntz},
  title        = {joshspeagle/dynesty: v3.0.0},
  month        = oct,
  year         = 2025,
  publisher    = {Zenodo},
  version      = {v3.0.0},
  doi          = {10.5281/zenodo.17268284},
  url          = {https://doi.org/10.5281/zenodo.17268284},
  swhid        = {swh:1:dir:b3340f6aaa931bae6c6bcff6e7ddeb89ca508a15
                   ;origin=https://doi.org/10.5281/zenodo.3348367;vis
                   it=swh:1:snp:fa92422c27d5611b107216e9b766a871bb43b
                   bee;anchor=swh:1:rel:94af3a8ef6e50d997258e152c3833
                   3550bd7463f;path=joshspeagle-dynesty-217bc94
                  },
}

@article{dotter_influence_2017,
	title = {The {Influence} of {Atomic} {Diffusion} on {Stellar} {Ages} and {Chemical} {Tagging}},
	volume = {840},
	issn = {0004-637X},
	url = {https://doi.org/10.3847/1538-4357/aa6d10},
	doi = {10.3847/1538-4357/aa6d10},
	language = {en},
	number = {2},
	urldate = {2026-02-01},
	journal = {The Astrophysical Journal},
	publisher = {The American Astronomical Society},
	author = {Dotter, Aaron and Conroy, Charlie and Cargile, Phillip and Asplund, Martin},
	month = may,
	year = {2017},
	pages = {99},
}

@article{moedas_atomic_2022,
	title = {Atomic diffusion and turbulent mixing in solar-like stars: {Impact} on the fundamental properties of {FG}-type stars},
	volume = {666},
	copyright = {© N. Moedas et al. 2022},
	issn = {0004-6361, 1432-0746},
	shorttitle = {Atomic diffusion and turbulent mixing in solar-like stars},
	url = {https://www.aanda.org/articles/aa/abs/2022/10/aa43210-22/aa43210-22.html},
	doi = {10.1051/0004-6361/202243210},
	language = {en},
	urldate = {2026-06-22},
	journal = {Astronomy \& Astrophysics},
	publisher = {EDP Sciences},
	author = {Moedas, Nuno and Deal, Morgan and Bossini, Diego and Campilho, Bernardo},
	month = oct,
	year = {2022},
	pages = {A43},
}

@ARTICLE{Yang2023,
       author = {{Yang}, Jia-Yi and {Chen}, Di-Chang and {Xie}, Ji-Wei and {Zhou}, Ji-Lin and {Dong}, Subo and {Zhu}, Zi and {Zheng}, Zheng and {Liu}, Chao and {Zong}, Weikai and {Luo}, Ali},
        title = "{Planets Across Space and Time (PAST). IV. The Occurrence and Architecture of Kepler Planetary Systems as a Function of Kinematic Age Revealed by the LAMOST-Gaia-Kepler Sample}",
      journal = {\aj},
         year = 2023,
        month = dec,
       volume = {166},
       number = {6},
          eid = {243},
        pages = {243},
          doi = {10.3847/1538-3881/ad0368},
archivePrefix = {arXiv},
       eprint = {2310.20113},
 primaryClass = {astro-ph.EP},
       adsurl = {https://ui.adsabs.harvard.edu/abs/2023AJ....166..243Y}
}

@article{Spaargaren2025,
   author = {Rob J. Spaargaren and Oliver Herbort and Haiyang S. Wang and Stephen J. Mojzsis and Paolo Sossi},
   doi = {10.1051/0004-6361/202556011},
   issn = {14320746},
   journal = {Astronomy and Astrophysics},
   month = {11},
   publisher = {EDP Sciences},
   title = {Proto-planetary disk composition-dependent element volatility in the context of rocky planet formation},
   volume = {703},
   year = {2025}
}

@article{Zaveri2026,
   author = {Urja Zaveri and Haiyang S. Wang and Paolo A. Sossi},
   doi = {10.1051/0004-6361/202558126},
   issn = {0004-6361},
   journal = {Astronomy \& Astrophysics},
   month = {5},
   pages = {A223},
   title = {A chemical perspective on planet formation in reduced systems},
   volume = {709},
   url = {https://www.aanda.org/10.1051/0004-6361/202558126},
   year = {2026}
}

@INPROCEEDINGS{Hayashi1985,
       author = {{Hayashi}, C. and {Nakazawa}, K. and {Nakagawa}, Y.},
        title = "{Formation of the solar system.}",
    booktitle = {Protostars and Planets II},
         year = 1985,
       editor = {{Black}, D.~C. and {Matthews}, M.~S.},
        month = jan,
        pages = {1100-1153},
       adsurl = {https://ui.adsabs.harvard.edu/abs/1985prpl.conf.1100H}
}

@ARTICLE{Laskar2009,
       author = {{Laskar}, J. and {Gastineau}, M.},
        title = "{Existence of collisional trajectories of Mercury, Mars and Venus with the Earth}",
      journal = {\nat},
         year = 2009,
        month = jun,
       volume = {459},
       number = {7248},
        pages = {817-819},
          doi = {10.1038/nature08096},
       adsurl = {https://ui.adsabs.harvard.edu/abs/2009Natur.459..817L}
}

@ARTICLE{bedell_abundance_2014,
       author = {{Bedell}, Megan and {Mel{\'e}ndez}, Jorge and {Bean}, Jacob L. and {Ram{\'\i}rez}, Ivan and {Leite}, Paulo and {Asplund}, Martin},
        title = "{Stellar Chemical Abundances: In Pursuit of the Highest Achievable Precision}",
      journal = {\apj},
         year = 2014,
        month = nov,
       volume = {795},
       number = {1},
          eid = {23},
        pages = {23},
          doi = {10.1088/0004-637X/795/1/23},
archivePrefix = {arXiv},
       eprint = {1409.1230},
 primaryClass = {astro-ph.SR},
       adsurl = {https://ui.adsabs.harvard.edu/abs/2014ApJ...795...23B}
}

@article{sun_planets_2025,
	title = {Planets {Around} {Solar} {Twins}/{Analogs} ({PASTA}): {II}. {Chemical} abundances, systematic offsets, and clues as to planet formation},
	volume = {701},
	issn = {0004-6361},
	shorttitle = {Planets {Around} {Solar} {Twins}/{Analogs} ({PASTA})},
	url = {https://ui.adsabs.harvard.edu/abs/2025A&A...701A.107S},
	doi = {10.1051/0004-6361/202556272},
	urldate = {2025-09-25},
	journal = {Astronomy and Astrophysics},
	publisher = {EDP},
	author = {Sun, Qinghui and Ji, Chenyang and Wang, Sharon Xuesong and Lin, Zitao and Teske, Johanna and Ting, Yuan-Sen and Bedell, Megan and Liu, Fan},
	month = sep,
	year = {2025},
	note = {ADS Bibcode: 2025A\&A...701A.107S},
	pages = {A107},
}

@article{melendez_peculiar_2009,
	title = {The {Peculiar} {Solar} {Composition} and {Its} {Possible} {Relation} to {Planet} {Formation}},
	volume = {704},
	issn = {0004-637X},
	url = {https://ui.adsabs.harvard.edu/abs/2009ApJ...704L..66M},
	doi = {10.1088/0004-637X/704/1/L66},
	urldate = {2025-09-25},
	journal = {The Astrophysical Journal},
	author = {Meléndez, J. and Asplund, M. and Gustafsson, B. and Yong, D.},
	month = oct,
	year = {2009},
	note = {ADS Bibcode: 2009ApJ...704L..66M},
	pages = {L66--L70},
}

@article{bedell_chemical_2018,
	title = {The {Chemical} {Homogeneity} of {Sun}-like {Stars} in the {Solar} {Neighborhood}},
	volume = {865},
	issn = {0004-637X},
	url = {https://ui.adsabs.harvard.edu/abs/2018ApJ...865...68B},
	doi = {10.3847/1538-4357/aad908},
	urldate = {2025-09-25},
	journal = {The Astrophysical Journal},
	author = {Bedell, Megan and Bean, Jacob L. and Meléndez, Jorge and Spina, Lorenzo and Ramírez, Ivan and Asplund, Martin and Alves-Brito, Alan and dos Santos, Leonardo and Dreizler, Stefan and Yong, David and Monroe, TalaWanda and Casagrande, Luca},
	month = sep,
	year = {2018},
	note = {ADS Bibcode: 2018ApJ...865...68B},
	pages = {68},
}

@article{liu_at_2024,
	title = {At least one in a dozen stars shows evidence of planetary ingestion},
	volume = {627},
	issn = {0028-0836},
	url = {https://ui.adsabs.harvard.edu/abs/2024Natur.627..501L},
	doi = {10.1038/s41586-024-07091-y},
	urldate = {2025-09-25},
	journal = {Nature},
	author = {Liu, Fan and Ting, Yuan-Sen and Yong, David and Bitsch, Bertram and Karakas, Amanda and Murphy, Michael T. and Joyce, Meridith and Dotter, Aaron and Dai, Fei},
	month = mar,
	year = {2024},
	note = {ADS Bibcode: 2024Natur.627..501L},
	pages = {501--504},
}

@article{sun_planets_2025-1,
	title = {Planets {Around} {Solar} {Twins}/{Analogs} ({PASTA}). {I}. {High}-precision {Stellar} {Chemical} {Abundances} for 17 {Planet}-hosting {Stars} and the {Condensation} {Temperature} {Trend}},
	volume = {980},
	issn = {0004-637X},
	url = {https://ui.adsabs.harvard.edu/abs/2025ApJ...980..179S},
	doi = {10.3847/1538-4357/ad9924},
	urldate = {2025-09-25},
	journal = {The Astrophysical Journal},
	publisher = {IOP},
	author = {Sun, Qinghui and Wang, Sharon Xuesong and Gan, Tianjun and Ji, Chenyang and Lin, Zitao and Ting, Yuan-Sen and Teske, Johanna and Li, Haining and Liu, Fan and Hua, Xinyan and Tang, Jiaxin and Yu, Jie and Zhang, Jiayue and Badenas-Agusti, Mariona and Vanderburg, Andrew and Ricker, George R. and Vanderspek, Roland and Latham, David W. and Seager, Sara and Jenkins, Jon M. and Schwarz, Richard P. and Guillot, Tristan and Tan, Thiam-Guan and Conti, Dennis M. and Collins, Kevin I. and Srdoc, Gregor and Stockdale, Chris and Suarez, Olga and Zambelli, Roberto and Radford, Don and Barkaoui, Khalid and Evans, Phil and Bieryla, Allyson},
	month = feb,
	year = {2025},
	note = {ADS Bibcode: 2025ApJ...980..179S},
	pages = {179},
}

@article{spina_temporal_2018,
	title = {The temporal evolution of neutron-capture elements in the {Galactic} discs},
	volume = {474},
	issn = {0035-8711},
	url = {https://ui.adsabs.harvard.edu/abs/2018MNRAS.474.2580S},
	doi = {10.1093/mnras/stx2938},
	urldate = {2025-10-03},
	journal = {Monthly Notices of the Royal Astronomical Society},
	publisher = {OUP},
	author = {Spina, Lorenzo and Meléndez, Jorge and Karakas, Amanda I. and dos Santos, Leonardo and Bedell, Megan and Asplund, Martin and Ramírez, Ivan and Yong, David and Alves-Brito, Alan and Bean, Jacob L. and Dreizler, Stefan},
	month = feb,
	year = {2018},
	note = {ADS Bibcode: 2018MNRAS.474.2580S},
	pages = {2580--2593},
}

@article{rampalli_sun_2024,
	title = {The {Sun} {Remains} {Relatively} {Refractory} {Depleted}: {Elemental} {Abundances} for 17,412 {Gaia} {RVS} {Solar} {Analogs} and 50 {Planet} {Hosts}},
	volume = {965},
	issn = {0004-637X},
	shorttitle = {The {Sun} {Remains} {Relatively} {Refractory} {Depleted}},
	url = {https://ui.adsabs.harvard.edu/abs/2024ApJ...965..176R},
	doi = {10.3847/1538-4357/ad303e},
	urldate = {2025-10-18},
	journal = {The Astrophysical Journal},
	publisher = {IOP},
	author = {Rampalli, Rayna and Ness, Melissa K. and Edwards, Graham H. and Newton, Elisabeth R. and Bedell, Megan},
	month = apr,
	year = {2024},
	note = {ADS Bibcode: 2024ApJ...965..176R},
	pages = {176},
}

@article{asplund_chemical_2009,
	title = {The {Chemical} {Composition} of the {Sun}},
	volume = {47},
	issn = {0066-4146},
	url = {https://ui.adsabs.harvard.edu/abs/2009ARA&A..47..481A},
	doi = {10.1146/annurev.astro.46.060407.145222},
	urldate = {2025-10-26},
	journal = {Annual Review of Astronomy and Astrophysics},
	author = {Asplund, Martin and Grevesse, Nicolas and Sauval, A. Jacques and Scott, Pat},
	month = sep,
	year = {2009},
	note = {ADS Bibcode: 2009ARA\&A..47..481A},
	pages = {481--522},
}

@article{chambers_slar_2010,
	title = {S℡{LAR} {ELEMENTAL} {ABUNDANCE} {PATTERNS}: {IMPLICATIONS} {FOR} {PLANET} {FORMATION}},
	volume = {724},
	issn = {0004-637X},
	shorttitle = {S℡{LAR} {ELEMENTAL} {ABUNDANCE} {PATTERNS}},
	url = {https://doi.org/10.1088/0004-637X/724/1/92},
	doi = {10.1088/0004-637X/724/1/92},
	language = {en},
	number = {1},
	urldate = {2026-02-01},
	journal = {The Astrophysical Journal},
	publisher = {The American Astronomical Society},
	author = {Chambers, J. E.},
	month = oct,
	year = {2010},
	pages = {92},
}

@article{kunitomo_revisiting_2018,
	title = {Revisiting the pre-main-sequence evolution of stars - {II}. {Consequences} of planet formation on stellar surface composition},
	volume = {618},
	copyright = {© ESO 2018},
	issn = {0004-6361, 1432-0746},
	url = {https://www.aanda.org/articles/aa/abs/2018/10/aa33127-18/aa33127-18.html},
	doi = {10.1051/0004-6361/201833127},
	language = {en},
	urldate = {2026-02-01},
	journal = {Astronomy \& Astrophysics},
	publisher = {EDP Sciences},
	author = {Kunitomo, Masanobu and Guillot, Tristan and Ida, Shigeru and Takeuchi, Taku},
	month = oct,
	year = {2018},
	pages = {A132},
}

@article{adibekyan_origin_2014,
	title = {On the origin of stars with and without planets - {Tc} trends and clues to {Galactic} evolution},
	volume = {564},
	copyright = {© ESO, 2014},
	issn = {0004-6361, 1432-0746},
	url = {https://www.aanda.org/articles/aa/abs/2014/04/aa23435-14/aa23435-14.html},
	doi = {10.1051/0004-6361/201423435},
	language = {en},
	urldate = {2026-02-01},
	journal = {Astronomy \& Astrophysics},
	publisher = {EDP Sciences},
	author = {Adibekyan, V. Zh and Hernández, J. I. González and Mena, E. Delgado and Sousa, S. G. and Santos, N. C. and Israelian, G. and Figueira, P. and Lis, S. Bertran de},
	month = apr,
	year = {2014},
	pages = {L15},
}

@article{nissen_high-precision_2015,
	title = {High-precision abundances of elements in solar twin stars - {Trends} with stellar age and elemental condensation temperature},
	volume = {579},
	copyright = {© ESO, 2015},
	issn = {0004-6361, 1432-0746},
	url = {https://www.aanda.org/articles/aa/abs/2015/07/aa26269-15/aa26269-15.html},
	doi = {10.1051/0004-6361/201526269},
	language = {en},
	urldate = {2026-02-01},
	journal = {Astronomy \& Astrophysics},
	publisher = {EDP Sciences},
	author = {Nissen, P. E.},
	month = jul,
	year = {2015},
	pages = {A52},
}

@article{booth_fingerprints_2020,
	title = {Fingerprints of giant planets in the composition of solar twins},
	volume = {493},
	issn = {0035-8711},
	url = {https://doi.org/10.1093/mnras/staa578},
	doi = {10.1093/mnras/staa578},
	number = {4},
	urldate = {2026-02-01},
	journal = {Monthly Notices of the Royal Astronomical Society},
	author = {Booth, Richard A and Owen, James E},
	month = apr,
	year = {2020},
	pages = {5079--5088},
}

@article{spina_gaia-eso_2015,
	title = {The {Gaia}-{ESO} {Survey}: chemical signatures of rocky accretion in a young solar-type star},
	volume = {582},
	copyright = {© ESO, 2015},
	issn = {0004-6361, 1432-0746},
	shorttitle = {The {Gaia}-{ESO} {Survey}},
	url = {https://www.aanda.org/articles/aa/abs/2015/10/aa26896-15/aa26896-15.html},
	doi = {10.1051/0004-6361/201526896},
	language = {en},
	urldate = {2026-02-01},
	journal = {Astronomy \& Astrophysics},
	publisher = {EDP Sciences},
	author = {Spina, L. and Palla, F. and Randich, S. and Sacco, G. and Jeffries, R. and Magrini, L. and Franciosini, E. and Meyer, M. R. and Tautvaišienė, G. and Gilmore, G. and Alfaro, E. J. and Prieto, C. Allende and Bensby, T. and Bragaglia, A. and Flaccomio, E. and Koposov, S. E. and Lanzafame, A. C. and Costado, M. T. and Hourihane, A. and Lardo, C. and Lewis, J. and Monaco, L. and Morbidelli, L. and Sousa, S. G. and Worley, C. C. and Zaggia, S.},
	month = oct,
	year = {2015},
	pages = {L6},
}

@article{oh_kronos_2018,
	title = {Kronos and {Krios}: {Evidence} for {Accretion} of a {Massive}, {Rocky} {Planetary} {System} in a {Comoving} {Pair} of {Solar}-type {Stars}},
	volume = {854},
	issn = {0004-637X},
	shorttitle = {Kronos and {Krios}},
	url = {https://doi.org/10.3847/1538-4357/aaab4d},
	doi = {10.3847/1538-4357/aaab4d},
	language = {en},
	number = {2},
	urldate = {2026-02-01},
	journal = {The Astrophysical Journal},
	publisher = {The American Astronomical Society},
	author = {Oh, Semyeong and Price-Whelan, Adrian M. and Brewer, John M. and Hogg, David W. and Spergel, David N. and Myles, Justin},
	month = feb,
	year = {2018},
	pages = {138},
}

@article{weinberg_chemical_2022,
	title = {Chemical {Cartography} with {APOGEE}: {Mapping} {Disk} {Populations} with a 2-process {Model} and {Residual} {Abundances}},
	volume = {260},
	issn = {0067-0049},
	shorttitle = {Chemical {Cartography} with {APOGEE}},
	url = {https://doi.org/10.3847/1538-4365/ac6028},
	doi = {10.3847/1538-4365/ac6028},
	language = {en},
	number = {2},
	urldate = {2026-02-01},
	journal = {The Astrophysical Journal Supplement Series},
	publisher = {The American Astronomical Society},
	author = {Weinberg, David H. and Holtzman, Jon A. and Johnson, Jennifer A. and Hayes, Christian and Hasselquist, Sten and Shetrone, Matthew and Ting, Yuan-Sen and Beaton, Rachael L. and Beers, Timothy C. and Bird, Jonathan C. and Bizyaev, Dmitry and Blanton, Michael R. and Cunha, Katia and Fernández-Trincado, José G. and Frinchaboy, Peter M. and García-Hernández, D. A. and Griffith, Emily and Johnson, James W. and Jönsson, Henrik and Lane, Richard R. and Leung, Henry W. and Mackereth, J. Ted and Majewski, Steven R. and Mészáros, Szabolcs and Nitschelm, Christian and Pan, Kaike and Schiavon, Ricardo P. and Schneider, Donald P. and Schultheis, Mathias and Smith, Verne and Sobeck, Jennifer S. and Stassun, Keivan G. and Stringfellow, Guy S. and Vincenzo, Fiorenzo and Wilson, John C. and Zasowski, Gail},
	month = jun,
	year = {2022},
	pages = {32},
}

@article{griffith_residual_2022,
	title = {Residual {Abundances} in {GALAH} {DR3}: {Implications} for {Nucleosynthesis} and {Identification} of {Unique} {Stellar} {Populations}},
	volume = {931},
	issn = {0004-637X},
	shorttitle = {Residual {Abundances} in {GALAH} {DR3}},
	url = {https://doi.org/10.3847/1538-4357/ac5826},
	doi = {10.3847/1538-4357/ac5826},
	language = {en},
	number = {1},
	urldate = {2026-02-01},
	journal = {The Astrophysical Journal},
	publisher = {The American Astronomical Society},
	author = {Griffith, Emily J. and Weinberg, David H. and Buder, Sven and Johnson, Jennifer A. and Johnson, James W. and Vincenzo, Fiorenzo},
	month = may,
	year = {2022},
	pages = {23},
}

@article{griffith_abundance_2019,
	title = {Abundance {Ratios} in {GALAH} {DR2} and {Their} {Implications} for {Nucleosynthesis}},
	volume = {886},
	issn = {0004-637X},
	url = {https://iopscience.iop.org/article/10.3847/1538-4357/ab4b5d},
	doi = {10.3847/1538-4357/ab4b5d},
	language = {en},
	number = {2},
	urldate = {2026-02-01},
	journal = {The Astrophysical Journal},
	publisher = {IOP Publishing},
	author = {Griffith, Emily and Johnson, Jennifer A. and Weinberg, David H.},
	month = nov,
	year = {2019},
	pages = {84},
}

@article{weinberg_chemical_2019,
	title = {Chemical {Cartography} with {APOGEE}: {Multi}-element {Abundance} {Ratios}},
	volume = {874},
	issn = {0004-637X},
	shorttitle = {Chemical {Cartography} with {APOGEE}},
	url = {https://iopscience.iop.org/article/10.3847/1538-4357/ab07c7},
	doi = {10.3847/1538-4357/ab07c7},
	language = {en},
	number = {1},
	urldate = {2026-02-01},
	journal = {The Astrophysical Journal},
	publisher = {IOP Publishing},
	author = {Weinberg, David H. and Holtzman, Jon A. and Hasselquist, Sten and Bird, Jonathan C. and Johnson, Jennifer A. and Shetrone, Matthew and Sobeck, Jennifer and Prieto, Carlos Allende and Bizyaev, Dmitry and Carrera, Ricardo and Cohen, Roger E. and Cunha, Katia and Ebelke, Garrett and Fernandez-Trincado, J. G. and García-Hernández, D. A. and Hayes, Christian R. and Jönsson, Henrik and Lane, Richard R. and Majewski, Steven R. and Malanushenko, Viktor and Mészáros, Szabolcs and Nidever, David L. and Nitschelm, Christian and Pan, Kaike and Rix, Hans-Walter and Rybizki, Jan and Schiavon, Ricardo P. and Schneider, Donald P. and Wilson, John C. and Zamora, Olga},
	month = mar,
	year = {2019},
	pages = {102},
}

@article{pinsonneault_mass_2001,
	title = {The {Mass} of the {Convective} {Zone} in {FGK} {Main}-{Sequence} {Stars} and the {Effect} of {Accreted} {Planetary} {Material} on {Apparent} {Metallicity} {Determinations}},
	volume = {556},
	issn = {0004-637X},
	url = {https://ui.adsabs.harvard.edu/abs/2001ApJ...556L..59P},
	doi = {10.1086/323531},
	urldate = {2026-02-02},
	journal = {The Astrophysical Journal},
	publisher = {IOP},
	author = {Pinsonneault, M. H. and DePoy, D. L. and Coffee, M.},
	month = jul,
	year = {2001},
	note = {ADS Bibcode: 2001ApJ...556L..59P},
	pages = {L59--L62},
}

@article{huhn_how_2023,
	title = {How accretion of planet-forming disks influences stellar abundances},
	volume = {676},
	issn = {0004-6361},
	url = {https://ui.adsabs.harvard.edu/abs/2023A&A...676A..87H},
	doi = {10.1051/0004-6361/202346604},
	urldate = {2026-02-02},
	journal = {Astronomy and Astrophysics},
	publisher = {EDP},
	author = {Hühn, L.-A. and Bitsch, B.},
	month = aug,
	year = {2023},
	note = {ADS Bibcode: 2023A\&A...676A..87H},
	pages = {A87},
}

@article{wasson_compositions_1988,
	title = {Compositions of chondrites},
	volume = {325},
	issn = {0080-4614},
	url = {https://doi.org/10.1098/rsta.1988.0066},
	doi = {10.1098/rsta.1988.0066},
	number = {1587},
	urldate = {2026-02-02},
	journal = {Philosophical Transactions of the Royal Society of London, Series A: Mathematical and Physical Sciences},
	author = {Wasson, J. T. and Kallemeyn, G. W.},
	editor = {Runcorn, Stanley Keith and Turner, Grenville and Woolfson, Michael Mark},
	month = jul,
	year = {1988},
	pages = {535--544},
}

@article{allegre_chemical_2001,
	title = {Chemical composition of the {Earth} and the volatility control on planetary genetics},
	volume = {185},
	issn = {0012-821X},
	url = {https://ui.adsabs.harvard.edu/abs/2001E&PSL.185...49A},
	doi = {10.1016/S0012-821X(00)00359-9},
	urldate = {2026-02-02},
	journal = {Earth and Planetary Science Letters},
	publisher = {Elsevier},
	author = {Allègre, Claude and Manhès, Gérard and Lewin, {\'E}ric},
	month = feb,
	year = {2001},
	note = {ADS Bibcode: 2001E\&PSL.185...49A},
	pages = {49--69},
}

@article{behmard_planet_2023,
	title = {Planet engulfment detections are rare according to observations and stellar modelling},
	volume = {521},
	issn = {0035-8711},
	url = {https://ui.adsabs.harvard.edu/abs/2023MNRAS.521.2969B},
	doi = {10.1093/mnras/stad745},
	urldate = {2026-02-02},
	journal = {Monthly Notices of the Royal Astronomical Society},
	publisher = {OUP},
	author = {Behmard, Aida and Dai, Fei and Brewer, John M. and Berger, Travis A. and Howard, Andrew W.},
	month = may,
	year = {2023},
	note = {ADS Bibcode: 2023MNRAS.521.2969B},
	pages = {2969--2987},
}

@article{spina_chemical_2021,
	title = {Chemical evidence for planetary ingestion in a quarter of {Sun}-like stars},
	volume = {5},
	issn = {2397-3366},
	url = {https://ui.adsabs.harvard.edu/abs/2021NatAs...5.1163S},
	doi = {10.1038/s41550-021-01451-8},
	urldate = {2026-02-03},
	journal = {Nature Astronomy},
	author = {Spina, Lorenzo and Sharma, Parth and Meléndez, Jorge and Bedell, Megan and Casey, Andrew R. and Carlos, Marília and Franciosini, Elena and Vallenari, Antonella},
	month = nov,
	year = {2021},
	note = {ADS Bibcode: 2021NatAs...5.1163S},
	pages = {1163--1169},
}

@article{speagle_dynesty_2020,
	title = {{DYNESTY}: a dynamic nested sampling package for estimating {Bayesian} posteriors and evidences},
	volume = {493},
	issn = {0035-8711},
	shorttitle = {{DYNESTY}},
	url = {https://ui.adsabs.harvard.edu/abs/2020MNRAS.493.3132S},
	doi = {10.1093/mnras/staa278},
	urldate = {2026-02-04},
	journal = {Monthly Notices of the Royal Astronomical Society},
	publisher = {OUP},
	author = {Speagle, Joshua S.},
	month = apr,
	year = {2020},
	note = {ADS Bibcode: 2020MNRAS.493.3132S},
	pages = {3132--3158},
}

@article{martos_signatures_2025,
	title = {Signatures of planets and {Galactic} subpopulations in solar analogs: {Precise} chemical abundances with neural networks},
	volume = {699},
	issn = {0004-6361},
	shorttitle = {Signatures of planets and {Galactic} subpopulations in solar analogs},
	url = {https://ui.adsabs.harvard.edu/abs/2025A&A...699A..46M},
	doi = {10.1051/0004-6361/202554675},
	urldate = {2026-02-28},
	journal = {Astronomy and Astrophysics},
	publisher = {EDP},
	author = {Martos, Giulia and Meléndez, Jorge and Spina, Lorenzo and Lucatello, Sara},
	month = jul,
	year = {2025},
	note = {ADS Bibcode: 2025A\&A...699A..46M},
	pages = {A46},
}

@ARTICLE{2013A&A...558A..33A,
       author = {{Astropy Collaboration} and {Robitaille}, Thomas P. and
         {Tollerud}, Erik J. and {Greenfield}, Perry and {Droettboom}, Michael and
         {Bray}, Erik and {Aldcroft}, Tom and {Davis}, Matt and
         {Ginsburg}, Adam and {Price-Whelan}, Adrian M. and
         {Kerzendorf}, Wolfgang E. and {Conley}, Alexander and {Crighton}, Neil and
         {Barbary}, Kyle and {Muna}, Demitri and {Ferguson}, Henry and
         {Grollier}, Fr{\'e}d{\'e}ric and {Parikh}, Madhura M. and
         {Nair}, Prasanth H. and {Unther}, Hans M. and {Deil}, Christoph and
         {Woillez}, Julien and {Conseil}, Simon and {Kramer}, Roban and
         {Turner}, James E.~H. and {Singer}, Leo and {Fox}, Ryan and
         {Weaver}, Benjamin A. and {Zabalza}, Victor and {Edwards}, Zachary I. and
         {Azalee Bostroem}, K. and {Burke}, D.~J. and {Casey}, Andrew R. and
         {Crawford}, Steven M. and {Dencheva}, Nadia and {Ely}, Justin and
         {Jenness}, Tim and {Labrie}, Kathleen and {Lim}, Pey Lian and
         {Pierfederici}, Francesco and {Pontzen}, Andrew and {Ptak}, Andy and
         {Refsdal}, Brian and {Servillat}, Mathieu and {Streicher}, Ole},
        title = "{Astropy: A community Python package for astronomy}",
      journal = {\aap},
         year = "2013",
        month = "Oct",
       volume = {558},
          eid = {A33},
        pages = {A33},
          doi = {10.1051/0004-6361/201322068},
archivePrefix = {arXiv},
       eprint = {1307.6212},
 primaryClass = {astro-ph.IM},
       adsurl = {https://ui.adsabs.harvard.edu/abs/2013A&A...558A..33A}
}

@article{collaboration_astropy_2022,
	title = {The {Astropy} {Project}: {Sustaining} and {Growing} a {Community}-oriented {Open}-source {Project} and the {Latest} {Major} {Release} (v5.0) of the {Core} {Package}},
	volume = {935},
	issn = {0004-637X},
	shorttitle = {The {Astropy} {Project}},
	url = {https://ui.adsabs.harvard.edu/abs/2022ApJ...935..167A/abstract},
	doi = {10.3847/1538-4357/ac7c74},
	language = {en},
	number = {2},
	urldate = {2026-05-20},
	journal = {The Astrophysical Journal, Volume 935, Issue 2, id.167, 20 pp.},
	author = {{Astropy Collaboration} and Price-Whelan, Adrian M. and Lim, Pey Lian and Earl, Nicholas and Starkman, Nathaniel and Bradley, Larry and Shupe, David L. and Patil, Aarya A. and Corrales, Lia and Brasseur, C. E. and Nöthe, Maximilian and Donath, Axel and Tollerud, Erik and Morris, Brett M. and Ginsburg, Adam and Vaher, Eero and Weaver, Benjamin A. and Tocknell, James and Jamieson, William and van Kerkwijk, Marten H. and Robitaille, Thomas P. and Merry, Bruce and Bachetti, Matteo and Günther, H. Moritz and Aldcroft, Thomas L. and Alvarado-Montes, Jaime A. and Archibald, Anne M. and Bódi, Attila and Bapat, Shreyas and Barentsen, Geert and Bazán, Juanjo and Biswas, Manish and Boquien, Médéric and Burke, D. J. and Cara, Daria and Cara, Mihai and Conroy, Kyle E. and Conseil, Simon and Craig, Matthew W. and Cross, Robert M. and Cruz, Kelle L. and D'Eugenio, Francesco and Dencheva, Nadia and Devillepoix, Hadrien A. R. and Dietrich, Jörg P. and Eigenbrot, Arthur Davis and Erben, Thomas and Ferreira, Leonardo and Foreman-Mackey, Daniel and Fox, Ryan and Freij, Nabil and Garg, Suyog and Geda, Robel and Glattly, Lauren and Gondhalekar, Yash and Gordon, Karl D. and Grant, David and Greenfield, Perry and Groener, Austen M. and Guest, Steve and Gurovich, Sebastian and Handberg, Rasmus and Hart, Akeem and Hatfield-Dodds, Zac and Homeier, Derek and Hosseinzadeh, Griffin and Jenness, Tim and Jones, Craig K. and Joseph, Prajwel and Kalmbach, J. Bryce and Karamehmetoglu, Emir and Kałuszyński, Mikołaj and Kelley, Michael S. P. and Kern, Nicholas and Kerzendorf, Wolfgang E. and Koch, Eric W. and Kulumani, Shankar and Lee, Antony and Ly, Chun and Ma, Zhiyuan and MacBride, Conor and Maljaars, Jakob M. and Muna, Demitri and Murphy, N. A. and Norman, Henrik and O'Steen, Richard and Oman, Kyle A. and Pacifici, Camilla and Pascual, Sergio and Pascual-Granado, J. and Patil, Rohit R. and Perren, Gabriel I. and Pickering, Timothy E. and Rastogi, Tanuj and Roulston, Benjamin R. and Ryan, Daniel F. and Rykoff, Eli S. and Sabater, Jose and Sakurikar, Parikshit and Salgado, Jesús and Sanghi, Aniket and Saunders, Nicholas and Savchenko, Volodymyr and Schwardt, Ludwig and Seifert-Eckert, Michael and Shih, Albert Y. and Jain, Anany Shrey and Shukla, Gyanendra and Sick, Jonathan and Simpson, Chris and Singanamalla, Sudheesh and Singer, Leo P. and Singhal, Jaladh and Sinha, Manodeep and Sipőcz, Brigitta M. and Spitler, Lee R. and Stansby, David and Streicher, Ole and Šumak, Jani and Swinbank, John D. and Taranu, Dan S. and Tewary, Nikita and Tremblay, Grant R. and de Val-Borro, Miguel and Van Kooten, Samuel J. and Vasović, Zlatan and Verma, Shresth and de Miranda Cardoso, José Vinícius and Williams, Peter K. G. and Wilson, Tom J. and Winkel, Benjamin and Wood-Vasey, W. M. and Xue, Rui and Yoachim, Peter and Zhang, Chen and Zonca, Andrea and Contributors, Astropy Project},
	month = aug,
	year = {2022},
	pages = {167},
}
\bibliographystyle{aasjournalv7}



\end{document}